\documentclass[final]{iacrtrans}
\usepackage[utf8]{inputenc}
\usepackage{multicol}
\usepackage{multirow}
\usepackage{subfig}
\usepackage[ruled]{algorithm}
\usepackage{algorithmic}
\usepackage{enumerate}
\newtheorem{Definition}{Definition}
\newtheorem{Property}{Property}
\newtheorem{Assumption}{Assumption}

\author{Zheng Li\inst{1} \and Xiaoyang Dong\inst{1,2*} \and Xiaoyun Wang\inst{1,2}\thanks{Corresponding authors}}
\institute{Key Laboratory of Cryptologic Technology and Information
Security, Ministry of Education, Shandong University, P. R. China, \\
 \email[lizhengcn@mail.sdu.edu.cn, dongxiaoyang@mail.sdu.edu.cn]{{lizhengcn, dongxiaoyang}@mail.sdu.edu.cn} \and Institute for Advanced Study, Tsinghua University, P. R. China, \email{xiaoyunwang@tsinghua.edu.cn}.}
\title[Conditional Cube Attack on Round-Reduced ASCON]{Conditional Cube Attack on Round-Reduced ASCON}

\begin{document}
\maketitle

\keywords[\publname, ToSC, LaTeX]{\publname \and ToSC \and \LaTeX}

\maketitle
\begin{abstract}
This paper evaluates the secure level of authenticated encryption \textsc{Ascon} against cube-like method. \textsc{Ascon} submitted by Dobraunig \emph{et~al.} is one of 16 survivors of the 3rd round  CAESAR competition. The cube-like method is first used by Dinur \emph{et~al.} to analyze Keccak keyed modes. At CT-RSA 2015, Dobraunig \emph{et~al.} applied this method to 5/6-round reduced \textsc{Ascon}, whose structure is similar to Keccak keyed modes. However, for \textsc{Ascon} the non-linear layer
is more complex  and state is much smaller, which make it hard for the attackers to select enough cube variables that do not multiply with each other after the first round. This seems to be the reason why the best previous key-recovery attack is on 6-round \textsc{Ascon}, while for Keccak keyed modes (Keccak-MAC and Keyak) the attacked round is no less than 7-round.

In this paper, we generalize the conditional cube attack proposed by Huang \emph{et~al.}, and find new cubes depending on some key bit conditions for 5/6-round reduced \textsc{Ascon}, and translate the previous theoretic 6-round attack with $2^{66}$ time complexity to a practical one with $2^{40}$ time complexity. Moreover, we propose the first 7-round key-recovery attack on \textsc{Ascon}. By introducing \emph{the cube-like key-subset technique}, we divide the full key space into many subsets according to different key conditions. For each key subset, we launch the cube tester to determine if the key falls into it. Finally, we recover the full key space by testing all the key subsets. The total time complexity is about $2^{103.9}$. In addition, for a weak-key subset, whose size is $2^{117}$, the attack is more efficient and costs only $2^{77}$ time complexity. Those attacks do not threaten the full round (12 rounds) \textsc{Ascon}.
\end{abstract}

\section*{Introduction}
Nowadays, when confidential messages are transmitted using an insecure channel, both their privacy and integrity are usually needed. Authenticated encryption (AE) schemes are proposed to meet both goals simultaneously. In 2014, the CAESAR competition \cite{CAESAR} was launched to identify good authenticated encryption (AE) candidates as better alternatives to current options such as AES-GCM \cite{AES-GCM}. Totally, 57 candidates have been submitted to the first round of the CAESAR competition. After two rounds of assessments from world-wide cryptographers and engineers, only 16 survivors were announced to be included in the third round of the CAESAR competition in 15 Aug 2016.
In order to get a secure scheme, many more cryptographic analyses on these candidates are needed urgently.

\textsc{Ascon} \cite{ASCON} is one of the 16 candidates, which is submitted by Dobraunig \emph{et~al.} \textsc{Ascon} uses a lightweight  sponge construction, the internal state is only 320-bit. The previous best attack \cite{DBLP:conf/ctrsa/DobraunigEMS15} is proposed by \textsc{Ascon}'s authors themselves, where the key-recovery attack is on 6 out of 12-round  using a cube-like method with time complexity of $2^{66}$. In this paper, we will focus on the analysis of \textsc{Ascon} against cube-like method.

Cube attack \cite{DBLP:conf/eurocrypt/DinurS09} is a chosen plaintext key-recovery attack, which was introduced by Dinur and Shamir. Since then, cube attack was applied to many different cryptographic primitives \cite{DBLP:conf/fse/AumassonDMS09,DBLP:conf/fse/DinurS11a,DBLP:conf/fse/FouqueV13}. At Eurocrypt 2015, Dinur \emph{et~al.} \cite{DBLP:conf/eurocrypt/DinurMPSS15} published a key-recovery attack on Keccak keyed modes, where the cube variables are selected not to multiply with each other after the first round, then the output degree of the polynomials is reduced. Later it was applied to \textsc{Ketje} \cite{ketje} by Dong \emph{et~al.} \cite{cryptoeprint:2017:159}. Huang \emph{et~al.} \cite{cryptoeprint:2016:790} proposed a new \emph{conditional cube attack} on Keccak keyed modes and presented an 8-round attack on Keyak. By restraining some bit conditions of the key, they obtain a new set of cube variables which not only do not multiply with each other after the first round, but also contains one cube variable that does not multiply with other cube variables after the second round, and then the output degree over cube variables is further reduced.

\subsection*{Our Contributions}

In this paper, we continue to explore the secure level of \textsc{Ascon} against cube-like method.
\textsc{Ascon} has a more complex non-linear layer and smaller state than Keccak keyed modes, which make it hard for the attackers to select enough cube variables for the 7-round attack that do not multiply with each other after the first round. This seems to be the reason why the best key-recovery attack is on 6-round \textsc{Ascon}, while for Keccak keyed modes (Keccak-MAC and Keyak) the attacked round is no less than 7-round. In this paper,
we firstly generalize the conditional cube attack, which is first proposed by Huang \emph{et~al.} \cite{cryptoeprint:2016:790}. By exploring the details of non-linear layer and the conditional cube attack method, we improve the complexity of the 6-round attack from $2^{66}$ to a practical one $2^{40}$. And then inspired by a so-called key-dependent strategy \cite{DBLP:conf/ctrsa/DongLJW15,DBLP:conf/fse/LiJWD15}, we develop a new \emph{cube-like key-subset technique}. Based on this technique, we construct a series of 65-dimension cubes for different key subsets, for each key subset we give a 7-round key-recovery attack, finally the key-recovery attacks cover the full key space. Our attacks work on the latest version \textsc{Ascon} v1.2. The results are summarized in Table \ref{tab:sumofattacks}.
Our contribution is four fold:

\begin{table}
\centering
\caption{Summary of Key-recovery Attacks on \textsc{Ascon}}\label{tab:sumofattacks}
\begin{tabular}{|c|c|c|c|}
\noalign{\smallskip}
\hline
Type & Attacked Rounds   & Time & Source \\
\hline
\multirow{2}{*}{Differential-Linear}& 4/12 & $2^{18}$ & \cite{DBLP:conf/ctrsa/DobraunigEMS15}\\

& 5/12 & $2^{36}$ & \cite{DBLP:conf/ctrsa/DobraunigEMS15}\\
\hline
\multirow{6}{*}{Cube-like Method}& 5/12 & $2^{35}$ & \cite{DBLP:conf/ctrsa/DobraunigEMS15}\\
& 6/12 & $2^{66}$ & \cite{DBLP:conf/ctrsa/DobraunigEMS15}\\
\cline{2-4}
& 5/12 & $2^{24}$ & Section~\ref{sec:5rASCON}\\
& 6/12 & $2^{40}$ & Section~\ref{sec:6rASCON}\\
& 7/12 & $2^{103.9}$& Section~\ref{sec:7rASCON}\\
& 7/12 & $2^{77}$ for $2^{117}$ keys & Section~\ref{sec:7rASCON}\\
\hline
\noalign{\smallskip}
\end{tabular}
\end{table}
\begin{enumerate}
  \item We generalize the conditional cube attack proposed by Huang \emph{et~al.} \cite{cryptoeprint:2016:790}, where a cube variable of certain cube (all the cube variables do not multiply with each other after the first round) does not multiply with other cube variables under some key bit conditions after the second round. In our generalized model, these key bits actually produce some common divisors of all the cube sums on the output bits. That means, if these divisors are zero by restraining conditions on these key bits, all the cube sums will be zero. Hence, using cube testers, one can test whether these divisors are zero or not, and then deduce the key bit conditions.
 \item When applying the generalized conditional cube attack to \textsc{Ascon}, we find some sets of cube variables for \textsc{Ascon}. In each set, the cube variables do not multiply with each other after the first round. Moreover, by assigning some bit conditions of the key, one cube variable does not multiply with the other cube variables after the second round. This reduces the cube dimension and leads to 5/6-round key-recovery attacks with time complexity of $2^{24}$ and $2^{40}$, respectively. This is the first practical key-recovery attack on 6-round \textsc{Ascon}.
  \item In our 7-round attack, different from previous cube-like attacks \cite{DBLP:conf/eurocrypt/DinurMPSS15,cryptoeprint:2016:790,DBLP:conf/ctrsa/DobraunigEMS15} which make the cube variables not multiply with each other in the first round, we let two of the cube variables multiply with each other to generate quadratic terms after the first round. In the second round, by restraining some bit conditions of the key and adding some auxiliary variables, the produced quadratic terms do not multiply with other monomials of cube variables. This solves the problem that there are no enough cube variables to launch a 7-round cube attack on \textsc{Ascon}.
  \item By introducing \emph{the cube-like key-subset technique}, we construct many new 65-dimension cubes on \textsc{Ascon}, whose cube sum after 7th round is zero when some key bit conditions are met. In other words, if we divide the full key space according to the key bit conditions into many key subsets, we can test the cube sum of different 65-dimension cubes to determine which subset the secret key falls into and then determine which key bit conditions the secret key meets.
    Finally, the full key space is divided into 63 key subsets and one small remaining set\footnote{The remaining set means the secret key does not fall into any of the 63 key subsets.}.
    The computations of cube sums of different 65-dimension cubes are repeated on these subsets until the right key is recovered. If the right key is not recovered, we assume that it is in the remaining subset and search it for the right one. The time complexity of the 7-round key-recovery attack on \textsc{Ascon} is  $2^{103.9}$.   Moreover, if the key falls into a weak-key set, whose size is $2^{117}$, the total complexity is reduced to $2^{77}$.
\end{enumerate}
\subsection*{Organization of the Paper}
Section \ref{sec:preliminaries} gives some notations, a brief description of \textsc{Ascon} cipher, some related properties of S-box and our attack assumptions. In Section \ref{sec:relatedwork}, we briefly describe the related works. Section \ref{sec:model} introduces models. Then, the new 5/6-round conditional cube attacks on \textsc{Ascon} are introduced in Section \ref{sec:5rASCON} and \ref{sec:6rASCON}. In Section \ref{sec:7rASCON}, some new 65-dimension cubes corresponding to some \emph{partial divisors} are given, then 7-round conditional cube attack is launched on \textsc{Ascon}. Section \ref{sec:ascon128a} gives a discussion on \textsc{Ascon}-128a and a previous version \textsc{Ascon} v1.1. At last, we conclude this paper in Section \ref{sec:conclusion}.

\section{Preliminaries}\label{sec:preliminaries}
In this section, we will give some notations, a brief description of \textsc{Ascon}, some related properties of S-box, together with our attack assumptions.
\subsection{Notations}
\label{subsec:notations}
\begin{tabular}{ll}
$S_{i}$             & ~~the intermediate state after $i$-round, \\
                    & ~~for example $S_{0.5}$ means the intermediate state after S-box in 1st round, \\
                    & ~~esp. $S_{0}$ means the initial state of \textsc{Ascon}\\
$S_{i}[j]$          & ~~the $j$th word of $S_{i}$, $0\leqslant j\leqslant4$
\end{tabular}

\begin{tabular}{ll}
$S_{i}[j][k]$       & ~~the $k$th bit of $S_{i}[j]$, $0\leqslant j\leqslant4$, $0\leqslant k\leqslant63$\\
$v_{i}$             & ~~the $i$th cube variable\\
$IV(i)$             & ~~the $i$th bit of $IV$, $0\leqslant i\leqslant63$\\
$K$                 & ~~128-bit key, $K=k_0||k_1$\\
$k_0(i)$            & ~~the $i$th bit of $k_0$, $k_0$ is placed in $S_{0}[1]$, $0\leqslant i\leqslant63$\\
$k_1(i)$            & ~~the $i$th bit of $k_1$, $k_1$ is placed in $S_{0}[2]$, $0\leqslant i\leqslant63$\\
$n(i)$              & ~~the $i$th bit of $S_{0}[4]$, $0\leqslant i\leqslant63$\\
\end{tabular}

\subsection{Brief Description of \textsc{Ascon}}\label{subsec:Ascondescription}
Authenticated encryption cipher \textsc{Ascon} is one of the 16 candidates in 3rd round CAESAR competition, whose mode of operation is based on MonkeyDuplex \cite{MonkeyDuplex}. We give a brief description of the latest version \textsc{Ascon} v1.2 \cite{ASCON} proposed for the 3rd round CAESAR competition. It operates on a 320-bit state (five 64-bit words $x_{0},...,x_{4}$) in a sponge-like construction. Parameters of two flavors, \textsc{Ascon}-128 and \textsc{Ascon}-128a, are summarized in Table~\ref{tab:paraascon}, while \textsc{Ascon}-128 is the primary recommendation by the designers. Readers can refer to \cite{ASCON} for more details.\\

\begin{table}
\centering\caption{Parameters Set for \textsc{Ascon}}
\label{tab:paraascon}
\begin{tabular}{|l|c|c|c|c|c|c|c|}
\hline
\multirow{2}{*}{name}&\multicolumn{4}{c|}{bit size}&\multicolumn{2}{c|}{rounds}&\multirow{2}{*}{$IV$ values}\\
\cline{2-7}
                     &key&nonce&tag&data block&$p^{a}$&$p^{b}$&\\
\hline
\textsc{Ascon}-128          &128&128&128&64&12&6&$0x80400c0600000000$\\
\textsc{Ascon}-128a      &128&128&128&128&12&8&$0x80800c0800000000$\\
\hline
\noalign{\smallskip}
\end{tabular}
\end{table}

\begin{figure}
\centering
\includegraphics[height=2.9cm]{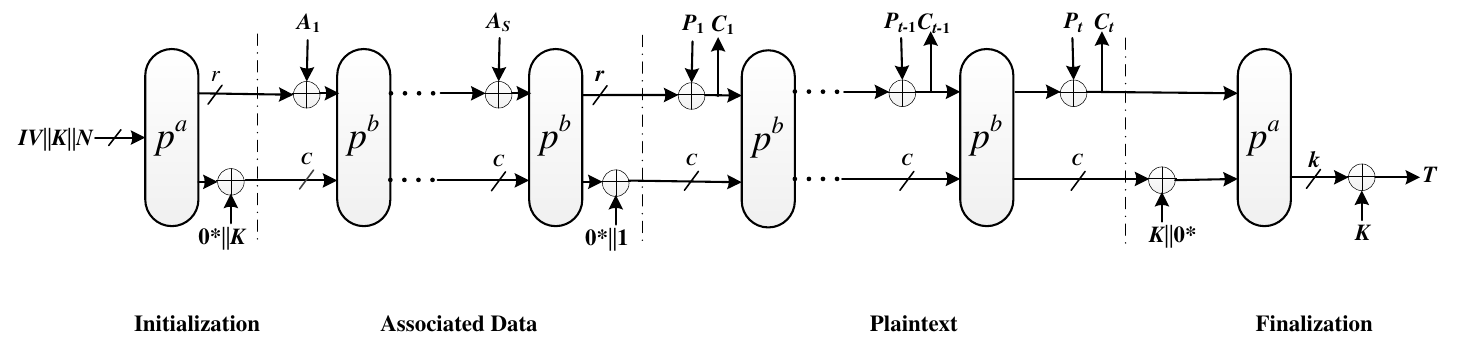}
\caption{The Encryption of \textsc{Ascon}}
\label{fig:ASCONwhole}
\end{figure}

\noindent\textbf{Mode.}~Based on MonkeyDuplex, the encryption of \textsc{Ascon} is organized in four phases as illustrated in Figure~\ref{fig:ASCONwhole}: initialization, processing associated data, processing the plaintext and finalization.
In the initialization, concatenation of $IV$(64-bit), the secret key $K$(128-bit) and nonce $N$(128-bit) initializes the state of \textsc{Ascon}, where $IV$ is constant in both flavors with values listed in Table~\ref{tab:paraascon}. It is processed by $p^{a}$ followed by XORing $K$. The associated 64-bit data block $A_{i}$ is XORed and then $p^{b}$ is applied to the intermediate state in sequence for $i=1,...,s$. A bit 1 is XORed to the least significant output bit of the last $p^{b}$ in the process of associated data. Each plaintext block $P_{i}$ with $i=1,...,t$ is processed similarly to $A_{i}$, while the corresponding $C_{i}$ is outputted. In the finalization, $K$ is XORed, then apply $p^{a}$. Finalization outputs $T$ after the $K$ is XORed to the least significant 128-bit.\\

\noindent\textbf{Permutation.}~Permutations $p^{a}$ and $p^{b}$ only differ in the number of iterations of round function $p$, which is shown in Table~\ref{tab:paraascon}. $p$ is composed of a constant addition to $x_{2}$, the substitution layer and the linear diffusion layer. Some constants are added in different rounds. The substitution layer applies a 5-bit S-box as shown in Table~\ref{tab:sboxascon} in parallel to each bit-slice of the five words $x_{0},x_{1},...,x_{4}$, where $x_{0}$ acts as the most significant bit of the S-box. It will be explored further in Section~\ref{sec:sbox}. The linear diffusion layer, shown in Eq.~(\ref{eq:linear}), provides diffusion in each 64-bit state-word $x_{i}$ with $\mathrm\Sigma_{i}(x_{i})$.

\makeatletter\def\@captype{table}\makeatother
\begin{center}
\begin{tabular}{ccccccccccccccccc}
\hline
$x$&0&1&2&3&4&5&6&7&8&9&10&11&12&13&14&15\\
\hline
$S(x)$&4&11&31&20&26&21&9&2&27&5&8&18&29&3&6&28\\
\hline
\hline
$x$&16&17&18&19&20&21&22&23&24&25&26&27&28&29&30&31\\
$S(x)$&30&19&7&14&0&13&17&24&16&12&1&25&22&10&15&23\\
\hline
\noalign{\smallskip}
\end{tabular}
\caption{The 5-bit S-box in the Substitution Layer of $p$}
\label{tab:sboxascon}
\end{center}

\begin{equation}\label{eq:linear}
\begin{aligned}
&\mathrm\Sigma_{0}(x_{0})=x_{0}\oplus (x_{0}\ggg19)\oplus (x_{0}\ggg28)\\
&\mathrm\Sigma_{1}(x_{1})=x_{1}\oplus (x_{1}\ggg61)\oplus (x_{1}\ggg39)\\
&\mathrm\Sigma_{2}(x_{2})=x_{2}\oplus (x_{2}\ggg~1)\oplus (x_{2}\ggg~6)\\
&\mathrm\Sigma_{3}(x_{3})=x_{3}\oplus (x_{3}\ggg10)\oplus (x_{3}\ggg17)\\
&\mathrm\Sigma_{4}(x_{4})=x_{4}\oplus (x_{4}\ggg~7)\oplus (x_{4}\ggg41)\\
\end{aligned}
\end{equation}

\subsection{Properties of S-box}\label{sec:sbox}
Denote the 5-bit input and output of the S-box as $x_0, x_1, x_2, x_3, x_4$ and $y_0, y_1, y_2, y_3, y_4$ respectively, and we use $x_0$ to mark the most significant bit or the first register word of the S-box. The algebraic normal form (ANF) of the S-box is as follow:
\[
\begin{aligned}
y_0 &= x_4x_1 + x_3 + x_2x_1 + x_2 + x_1x_0 + x_1 + x_0,      \\
y_1 &= x_4 + x_3x_2 + x_3x_1 + x_3 + x_2x_1 + x_2 + x_1 + x_0,\\
y_2 &= x_4x_3 + x_4 + x_2 + x_1 + 1,				           \\
y_3 &= x_4x_0 + x_4 + x_3x_0 + x_3 + x_2 + x_1 + x_0, \\
y_4 &= x_4x_1 + x_4 + x_3 + x_1x_0 + x_1.
\end{aligned}
\]

By studying the ANF of the S-box, the following two properties are given.
\begin{Property}\label{property:onlyx2}
Among the 5-bit output of the S-box, $x_4x_3$ only exists in $y_2$.
\end{Property}
If we choose $x_3$ and $x_4$ as cube variables, then the quadratic term $x_4x_3$ only exists in the ANF of $y_2$.

\begin{Property}\label{property:x2nox0x4}
$x_{2}$ will only multiply with $x_{1}$ and $x_{3}$. Especially, quadratic terms containing $x_2$ exist only in $y_0$ with $x_2x_1$ and $y_1$ with $x_3x_2+x_2x_1$.
\end{Property}
If $x_2$ is a cube variable and $x_1$ is a nonce bit, then we can select $x_1=0$ to delete $x_2x_1$.
\subsection{Our Attack Assumptions}\label{sec:attackassumption}
\begin{figure}
\centering
\includegraphics[height=2.9cm]{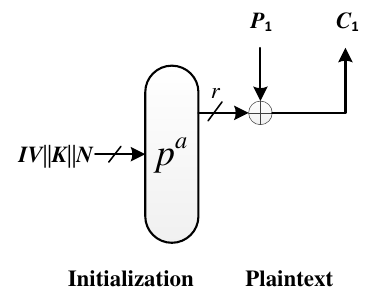}
\caption{Objective Procedure of \textsc{Ascon}}
\label{fig:ASCONpart}
\end{figure}

The \textsc{Ascon}'s design document \cite{ASCON} shows, if there is no associated data, the associated data processing phase will be removed. In our attack, some input and output bits of $p^{a}$ are needed, we omit the associated data processing phase too, i.e. our attack target is the initialization of \textsc{Ascon} as shown in Figure \ref{fig:ASCONpart}. This assumption is the same to \cite{DBLP:conf/ctrsa/DobraunigEMS15}. Our attacks work on both flavors \textsc{Ascon}-128 and \textsc{Ascon}-128a of the latest version \textsc{Ascon} v1.2, whose parameters are summarized in Table~\ref{tab:paraascon}. We describe the analysis of 5/6/7-round initialization of \textsc{Ascon}-128 in details, then give a discussion on \textsc{Ascon}-128a and \textsc{Ascon} v1.1 in Section~\ref{sec:ascon128a}.

\section{Related Work}\label{sec:relatedwork}
\subsection{Cube Attack}
The cube attack \cite{DBLP:conf/eurocrypt/DinurS09} was introduced by Dinur and Shamir at EUROCRYPT 2009.  It assumes that the output bit of a symmetric cryptographic scheme can be regarded as a polynomial $f(k_0,...,k_{n-1},v_0,...,v_{m-1})$ over $GF(2)$, $k_0,...,k_{n-1}$ are the secret variables (the key bits), $v_0,...,v_{m-1}$ are the public variables (e.g. $IV$ or nonce bits).

\begin{theorem}\label{the:cubetheorem}(\cite{DBLP:conf/eurocrypt/DinurS09})
\begin{equation}\label{eq.cubetest}
f(k_0,...,k_{n-1},v_0,...,v_{m-1}) = T \cdot {P} + {Q}(k_0,...,k_{n-1},v_0,...,v_{m-1})
\end{equation}
$T$ is a monomial which is actually the product of certain public variables, for example $(v_0,...,v_{s-1})$, $1\leq s\leq m$, denoted as cube $C_T$. None of the monomials in ${Q}$ is divisible by $T$. ${P}$ is called superpoly, which does not contain any variables of $C_T$. Then the sum of $f$ over all values of the cube $C_T$ (cube sum) is
\begin{equation}\label{eq.cubesum}
\sum\limits_{v'=(v_0,...,v_{s-1}) \in {C_T}} {f(k_0,...,k_{n-1},v',v_{s},...,v_{m-1}) = {P}}
\end{equation}
where $C_T$ contains all binary vectors of the length $s$, $v_{s},...,v_{m-1}$ are fixed to constant.
\end{theorem}
The basic idea is to find enough $T$ whose ${P}$ is linear and not a constant. This enables the key recovery through solving a system of linear equations.
\subsection{Dynamic Cube Attack}\label{subsec:dynamic}
Dynamic cube attack \cite{DBLP:conf/fse/DinurS11a} was first introduced to analyse Grain-128 by Dinur and Shamir at FSE 2011. The basic idea is to simplify a complex polynomial $P$:
$P=P_1\cdot P_2+P_3$
where $P_3$'s degree is relatively lower than $P$, and $P_1$ contains a linear public term called a dynamic variable. A dynamic variable is a variable assigned with a function in some secret variables (i.e. key bits) and cube variables to zero $P_1$. Thus, $P$ is simplified to $P_3$. One must firstly guess these key bits to compute dynamic variable. The right guess of key bits will lead to zero cube sums with high probability, otherwise the cube sums will be random.
\subsection{Conditional Differential Cryptanalysis}

Knellwolf, Meier and Naya-Plasenciaa \cite{DBLP:conf/asiacrypt/KnellwolfMN10} applied conditional differential characteristic
to NFSR-based constructions and extended to higher order differential attacks at ASIACRYPT 2010. The input of a synchronous stream cipher is an $IV$ and a key. Suppose that the keystream for many chosen $IV$s under the same secret key can be observed. By imposing specific conditions on certain bits of the $IV$, the attacker can control the propagation of a difference through the first few rounds of the initialization process. Taking $IV$ pairs conformed to these conditions as input, the resulting keystream differences will present a bias. Additionally, conditions upon key define classes of weak keys.

\section{Cube-like Attack Models}\label{sec:model}
In this section, we firstly generalize the \emph{conditional cube attack}, which was first introduced by Huang \emph{et~al.} to attack Keccak keyed mode. Then a new \emph{cube-like key-subset technique}  is introduced. At last, the rationality tests about these attack models are presented.

\subsection{Generalizing Conditional Cube Attack}\label{sec:conditionalcubeattack}
Conditional cube attack \cite{cryptoeprint:2016:790} was proposed by
Huang \emph{et~al.} to attack Keccak keyed mode. Inspired by dynamic cube attack \cite{DBLP:conf/eurocrypt/DinurS09}, which reduces the degree of output polynomials of cube variables by adding some bit conditions on $IV$, they reduce the degree by appending key bit conditions. The techniques are similar to message modification technique \cite{DBLP:conf/eurocrypt/WangY05,DBLP:conf/crypto/WangYY05} and conditional differential cryptanalysis \cite{DBLP:conf/asiacrypt/KnellwolfMN10} which used bit conditions to control differential propagation.
They also construct a cube tester based on cube variables and corresponding conditions, called \emph{conditional cube tester}.

In this section, we will generalize the conditional cube attack. In Theorem~\ref{the:cubetheorem}, the cube sum $P$ is calculated in Eq. (\ref{eq.cubesum}). In Keccak-MAC and other similar ciphers, after certain rounds, the cipher produces $l$-bit output. Each of the output bits is written as a polynomial $f_i(k_{0},...,k_{n-1},v_0,...,v_{m-1})$, $i = 0,1,...,l-1$. Choose a common cube  $C_T$, e.g $(v_{0},...,v_{s-1})$, $1\leq s\leq m$, then $f_i = T \cdot P_i  + {Q_i}$, $i = 0,1,...,l-1$. In conditional cube attack, a common divisor of $P_i$ is found, which is a polynomial $g(k_{0},...,k_{n-1},v_{s},...,v_{m-1})$. $v_{s},...,v_{m-1}$ are constant when computing cube sums $P_i=g(k_{0},...,k_{n-1},v_{s},...,v_{m-1}) \cdot {P'_i}$. Then the Corollary \ref{cor:cubecorollary} is given.

\begin{corollary}\label{cor:cubecorollary}
Given a series of polynomials $f_i$ $(i\in\{0,1,...,l-1\})$:\{0,1\}$^{n}\rightarrow$ \{0,1\}.
\begin{eqnarray}\label{eq.corcubetest}
\left\{
\begin{aligned}
&f_0(k_{0},...,k_{n-1},v_{0},...,v_{m-1}) = T \cdot g(k_{0},...,k_{n-1},v_{s},...,v_{m-1}) \cdot {P'_0} + {Q_0}\\
&f_1(k_{0},...,k_{n-1},v_{0},...,v_{m-1}) = T \cdot g(k_{0},...,k_{n-1},v_{s},...,v_{m-1}) \cdot {P'_1} + {Q_1}\\
&...\\
&f_{l-1}(k_{0},...,k_{n-1},v_{0},...,v_{m-1}) = T \cdot g(k_{0},...,k_{n-1},v_{s},...,v_{m-1}) \cdot {P'_{l-1}} + {Q_{l-1}}\\
\end{aligned}
\right.
\end{eqnarray}
where none of the monomials in ${Q_i}(x)$ is divisible by $T$. Then the sums of $f_i$ $(i\in\{0,1,...,l-1\})$ over all values of the cube (cube sum) are
\begin{eqnarray}\label{eq.corcubesum}
\left\{
\begin{aligned}
&\sum\limits_{v' \in {C_T}} {f_0(k_0,...,k_{n-1},v',v_{s},...,v_{m-1}) = g(k_{0},...,k_{n-1},v_{s},...,v_{m-1}) \cdot {P'_0}}\\
&\sum\limits_{v' \in {C_T}} {f_1(k_0,...,k_{n-1},v',v_{s},...,v_{m-1}) = g(k_{0},...,k_{n-1},v_{s},...,v_{m-1}) \cdot {P'_1}}\\
&...\\
&\sum\limits_{v' \in {C_T}} {f_{l-1}(k_0,...,k_{n-1},v',v_{s},...,v_{m-1}) = g(k_{0},...,k_{n-1},v_{s},...,v_{m-1}) \cdot {P'_{l-1}}}\\
\end{aligned}
\right.
\end{eqnarray}
where the $C_T$ contains all binary vectors of the length $s$.
\end{corollary}
Among the output polynomials, a common factor which is related to key instead of any cube bits exists. $g$ is introduced to represent the common factor for clearness.

As shown in Eq.~(\ref{eq.corcubesum}),
we get the following Property~\ref{pro:g0cf0} and Assumption~\ref{pro:g1cf1}.
\begin{Property}\label{pro:g0cf0}
If $g=0$, cube sums of $f_i~(i\in\{0,1,...,l-1\})$ will be all 0 with probability 1.
\end{Property}
\begin{Assumption}\label{pro:g1cf1}
If $g=1$, cube sums of $f_i~(i\in\{0,1,...,l-1\})$ will be determined by $P'_i~(i\in\{0,1,...,l-1\})$, the cube sums of $f_i~(i\in\{0,1,...,l-1\})$ all equal to 0 with probability about $2^{-l}$ if $f_i~(i\in\{0,1,...,l-1\})$ is a random oracle.
\end{Assumption}

According to Property~\ref{pro:g0cf0} and Assumption~\ref{pro:g1cf1}, we introduce the cube tester, which has the Property~\ref{pro:cubesumk0} and Assumption~\ref{pro:cubesumk1}.
\begin{Property}\label{pro:cubesumk0}
If at least one nonzero cube sum occurs among the cube sums of $f_i~(i\in\{0,1,...,l-1\})$, we will determine that $g=1$. It is guaranteed to be right.
\end{Property}
\begin{Assumption}\label{pro:cubesumk1}
If the cube sums of $f_i~(i\in\{0,1,...,l-1\})$ all equal to 0, we will determine that $g=0$. Note that, in a random oracle, $g=0$ is wrong with probability of $2^{-l}$, because $P'_i$ is zero with probability of about $\frac{1}{2}$.
\end{Assumption}

If the common divisor $g(k_{0},...,k_{n-1},v_{s},...,v_{m-1})$ is simple enough, e.g. $g=k_0$, we will use the cube tester of Property~\ref{pro:cubesumk0} and Assumption~\ref{pro:cubesumk1} to detect $k_0$.

\subsection{The Cube-like Key-subset Technique}\label{subsec:cubekey}

In Corollary \ref{cor:cubecorollary}, all cube sums ${P_i}=g(k_{0},...,k_{n-1},v_{s},...,v_{m-1}) \cdot {P'_i}$ have a common divisor $g$, hence $g=0$ will produce all zero cube sums illustrated in Property~\ref{pro:g0cf0}. When the $l$ cube sums $P_i$ do not have a common divisor, but the cube sums can be divided into several sets, there is a common divisor for cube sums in the same set, we denote these divisors as \emph{partial divisor}. For example, there are three divisors $g_1$, $g_2$ and $g_3$ which are the common divisor of cube sums $\{P_0,P_1,...,P_{19}\}$, $\{P_{20},P_{21},...,P_{39}\}$ and $\{P_{40},P_{41},...,P_{l-1}\}$, respectively. Then $g_1=0$, $g_2=0$ and  $g_3=0$ will produce all zero cube sums which is similar to Property~\ref{pro:g0cf0}.

Specially, the \emph{partial divisor}s $g_1$, $g_2$ and $g_3$ are some linear key-dependent polynomials, for example $g_1=k_1+a$, $g_2=k_2+b$ and $g_3=k_3+c$, where $k_1$, $k_2$ and $k_3$ are three key bit variables and $(a,b,c)$ are in $\mathbb{F}^3_2$.
If for each $(a,b,c)$ in $\mathbb{F}^3_2$, a cube (denoted as $C_{T(abc)}$) is found, that maintain the $g_1=k_1+a$, $g_2=k_2+b$ and $g_3=k_3+c$ to be \emph{partial divisor}s (which means  $g_1=0$, $g_2=0$ and $g_3=0$ will produce all zero cube sums). Then we can divide the full 128-bit key space (we assume that it is 128-bit) by $(a,b,c)$ to $2^{3}$ different key subsets. At last, we process the cube testers of all the $2^{3}$ cubes $C_{T(abc)}$ to determine which key subset the key candidate falls into. For example, according to Assumption~\ref{pro:cubesumk1}, if the cube tester of $C_{T(000)}$ produce all zero cube sums, we get $k_1=0$, $k_2=0$ and $k_3=0$. Hence the key candidates are reduced from the 128-bit full key space to a subset of size $2^{125}$, which meets the condition $k_1=0$, $k_2=0$ and $k_3=0$. We summarize those above techniques as \emph{cube-like key-subset technique}.

\subsection{Applications and Rationality test}\label{sec:rationality}

We apply the cube tester of Property~\ref{pro:cubesumk0} in our attacks on 5/6-round reduced initialization of \textsc{Ascon}. And we apply Assumption~\ref{pro:cubesumk1} and \emph{the cube-like key-subset technique} to 7-round reduced initialization of \textsc{Ascon}.
For \textsc{Ascon}-128, the number of output bits are 64-bit (128-bit for \textsc{Ascon}-128a). The success rate of the corresponding cube tester is dependent on the density of monomial $T$ in the output bits.

\begin{enumerate}
  \item If the monomial $T$ does not appear in anyone of $f_i$ $(i=0,1,...,63)$, the cube tester will fail to detect keys.
  \item If the monomial $T$ appears in some (a small fraction) of $f_i$ $(i=0,1,...,63)$,  we can use Property \ref{pro:cubesumk0} to detect the keys. When using Property \ref{pro:cubesumk0}, every detection returned is guaranteed to be right.
  \item If the monomial $T$ does appear in all (or most, e.g. $l_1$) of the $f_i$ $(i=0,1,...,63)$, we can use Property~\ref{pro:cubesumk0} or Assumption~\ref{pro:cubesumk1} to detect the keys. When using Assumption~\ref{pro:cubesumk1}, wrong key detection is returned with probability of about $2^{-64}$ or ${2^{-l_1}}$, besides, if there exist three partial divisors $g_1$, $g_2$ and $g_3$ as shown in Section~\ref{subsec:cubekey}, we get $g_1=0$, $g_2=0$ and $g_3=0$ with the same negligible probability of wrong detection.

\end{enumerate}

Thus, for a given cube, we should firstly test if the monomial $T$ appears in output $f_i$ $(i=0,1,...,63)$ of the round-reduced \textsc{Ascon}-128.
For 5/6-round attack on \textsc{Ascon} in Section~\ref{sec:5rASCON} and \ref{sec:6rASCON}, the monomial of the chosen cubes appears in the output $f_i$, we use Property~\ref{pro:cubesumk0} to detect the right key. The 5/6-round attacks are practical and the key-recovery programs are listed in \url{https://github.com/lizhengcn/Ascon_test}. But the 7-round attack is impractical, so in the following we present strong evidence for the correctness of our analysis. For 7-round attack, 65 (64+1) dimension cubes are chosen, so we have to test if the monomial $T$ of 65 degree appears in outputs $f_i$, however, the test should compute the cube sums of 65-dimension cube, which is out of reach by the computing resource. We test similar cubes that reduced to 17 (16+1) dimension for 5-round and 33 (32+1) dimension for 6-round, to deduce the appearance of monomial of 65 (64+1) dimension cubes. Both \textbf{Test 1} and \textbf{Test 2} provide evidences for the correctness of our assumptions.

\textbf{Test 1.}
We test 5-round cube sums of 1000 17-dimension random cubes:
For each of the 1000 simulations, we first place 16 cube variables in $S_0[3][i_j]~(j\in\{0,1,...,15\})$ randomly and place a cube variable in $S_0[4][i_0]$. The monomial (product of 17 variables) appears (that means the 64-bit cube sums of the 5-round are not all zero) in more than 850 simulations.

We test 6-round cube sum of 1000 33-dimension random cubes:
For each of the 1000 simulations, we firstly place 32 cube variables in $S_0[3][i_j]~(j\in\{0,1,...,31\})$ randomly, and place a cube variable in $S_0[4][i_0]$. The monomial (product of 33 variables) appears (that means the 64-bit cube sums of the 6-round are not all zero) in all the 1000 random tests.

So we deduce, in our attack for the 65-dimension cube (64 cube variables in $S_0[3][i_j]~(j\in\{0,1,...,63\})$, one cube variable in $S_0[4][i_0]$), the monomial (product of 65 variables) will also appear with significantly high probability.

\textbf{Test 2.}
We test 5-round cube sums over a 17-dimension random cube (cube variables are selected in the same way as Test 1.)  with 1000 keys. The monomial $T$ does really exist in some of $f_i$ $(i=0,1,...,63)$, as shown in Table \ref{tab:5r17dim1000keys}.

We test 6-round cube sums over a 33-dimension random cube with 982 keys. In each one of $f_i$ $(i=0,1,...,63)$, the cube sum is nonzero with a probability of about $\frac{1}{2}$, as shown in Table \ref{tab:6r33dim982keys}, which implies the monomial $T$ exists in all $f_i$s. Note that if the monomial $T$ does not exist in one of $f_i$s, e.g. $f_0$, the cube sum of $f_0$ is definitely zero. Experiments verify Assumption~\ref{pro:cubesumk1} for our 6-round attack on \textsc{Ascon}, so we conjecture it will also hold for the 7-round
attack.

The test programs are listed in \url{https://github.com/lizhengcn/Ascon_test}.

\section{Attack on 5-round initialization of \textsc{Ascon}}\label{sec:5rASCON}
\noindent\textbf{Basic Ideas.}
In the 5-round attack, we select a set of 16 cube variables $\{v_{0},v_{1}... v_{15}\}$. $\{v_i\}_{i=0,1,\ldots,15}$ are located in distinct S-boxes, thus they do not multiply with each other in the first round. With some bit conditions on key and nonce, $v_0$ does not multiply with any of $\{v_{1},v_{2}... v_{15}\}$ in the second round. Therefore, no $v_0v_i$ exists in $S_2$. As the algebraic degree of the round function is 2, $v_{0}v_{1}... v_{15}$ will not appear in $S_5$, which means the cube sums of the output of 5-round \textsc{Ascon} over $v_{0},v_{1}... v_{15}$ are all zero. If some of the cube sums are nonzero, bit conditions in the above are not met. According to Corollary \ref{cor:cubecorollary}, the bit conditions are actually determined by the common divisor of the cube sums.

In details, we construct a 16-dimension cube whose 64-bit cube sums have a common divisor $g=k_0(0)$. Similarly, we get $64\times 4=256$ 16-dimension cubes corresponding to different common divisors: $k_0(t),k_0(t)+1,k_0(t)+ k_1(t),k_0(t)+ k_1(t)+1$ with $t\in\{0,1,...,63\}$.
Thus, we apply Property \ref{pro:cubesumk0} to recover the key bits one by one. The cubes are listed in Table \ref{tab:5rattack} in Appendix \ref{app:parameters56}.\\
\begin{figure}
\centering
\includegraphics[height=2.4cm]{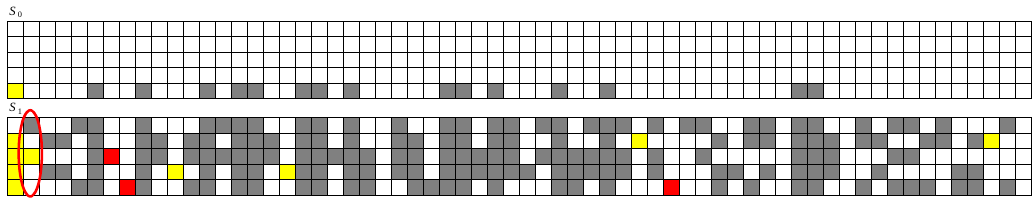}
\caption{Diffusion of $v_i$ with $k_0(0)=0$}
\label{fig:K[0][0]=0}
\end{figure}

\begin{figure}
\centering
\includegraphics[height=2.43cm]{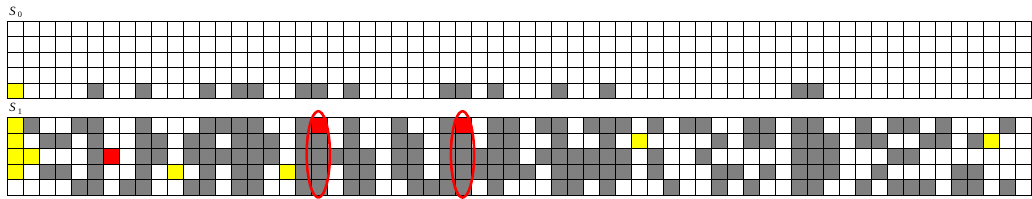}
\caption{Diffusion of $v_i$ with $k_0(0)=1$}
\label{fig:K[0][0]=1}
\end{figure}
\noindent\textbf{An Example to Determine $\bf{k_0(0)=1}$.}
For example, we select cube variables when $t=0$ and set nonce according to (1) of Table~\ref{tab:5rattack}. Figure~\ref{fig:K[0][0]=0} and~\ref{fig:K[0][0]=1} shows the diffusion of $v_i$ in the first round with $k_0(0)=0$ and $k_0(0)=1$ respectively, in which yellow bits represent the diffusion of $v_0$, and grey ones represent the diffusion of \{$v_{1},v_{2}\ldots v_{15}$\}, while a red one's polynomial is in form $h_1v_0+h_2v_i+h_3$ ($i\in\{1,\ldots,15\}, h_1,h_2,h_3$ only depend on nonce and key).

If $k_0(0)=0$, state $S_1$ of Figure~\ref{fig:K[0][0]=0} shows the distribution of $v_i$ at the input of the S-box layer in the second round: yellow and red bits are located in different S-boxes from grey ones except for $S_1[0][1]$ and $S_1[2][1]$ highlighted in a red circle. And Property~\ref{property:x2nox0x4} tells us that $S_1[0][1]$ does not multiply with $S_1[2][1]$. So $v_0$ will not multiply with any of $\{v_i\}_{i=1,2,\ldots,15}$ in the S-box operation of the second round and no $v_0v_i$ exists in $S_2$. As the algebraic degree of the round function is 2, a term in $S_5$ with the highest degree is product of 8 terms in $S_2$. If this term contains $v_0$ as a factor, its degree over cube variables $\{v_i\}_{i=0,1,\ldots,15}$ is at most 15 with all the other 7 terms being quadratic. If this term does not contain $v_0$, its degree over cube variables $\{v_i\}_{i=0,1,\ldots,15}$ will not exceed 15 obviously. Thus, $v_{0}v_{1}... v_{15}$ does not appear in $S_5$.

However, if $k_0(0)=1$, $v_0$ is certain to multiply with some of $\{v_i\}_{i=1,2,\ldots,15}$ as highlighted in red circles in Figure~\ref{fig:K[0][0]=1}. $v_0v_i$ exists in $S_2$. $v_{0}v_{1}... v_{15}$ will appear in $S_5$ with a high probability.
It means that $g=k_0(0)$ in Corollary \ref{cor:cubecorollary}. Therefore, if we obtain a nonzero cube sum, we can determine that $k_0(0)=1$, which is identical to Property \ref{pro:cubesumk0}.\\

\noindent\textbf{The Whole Procedure to Recover the Full Key.}
With parameters in Table~\ref{tab:5rattack}, 1-values for $k_0(t),k_0(t)+1,k_0(t)+ k_1(t),k_0(t)+ k_1(t)+1$ with $t\in\{0,1,...,63\}$ are detected by Property \ref{pro:cubesumk0}. Therefore, 0/1-values for $k_0(t),k_0(t)+ k_1(t)$ with $t\in\{0,1,...,63\}$ are detected. We describe the procedure in Algorithm~\ref{alg:testor56}.

We illustrate some special bit positions here. As a constant $0x000000000000000000f0$ is added to $S_0[2]$ before S-box in the first round, in which only 4 bits are 1 indexed by 56,57,58,59, $k_0(t)+ k_1(t)$ should reverse when $t\in\{56,57,58,59\}$. In \textsc{Ascon}-128 only six bits of $IV$ value are 1, whose indexes are $\{0,9,20,21,29,30\}$, and other $IV$ bits are 0, it leads to a special cube selection (not rotated from others) to detect 1-value of $k_0(t)+ k_1(t)+1$ for $t\in\{0,9,20,21,29,30\}$.

The whole procedure is performed as follow: with parameters in Table~\ref{tab:5rattack}, for $t\in\{0,1,\ldots,63\}$, if we obtain a nonzero cube sum in (1), we can determine that $k_0(t)=1$, if we obtain a nonzero cube sum in (2), we can determine that $k_0(t)=0$. For $t\in\{0,1,\ldots,63\}$ while $t\neq56,57,58,59$, if we obtain a nonzero cube sum in (3), we can determine that $k_0(t)+ k_1(t)= 1$; for $t\in\{56,57,58,59\}$, if we obtain a nonzero cube sum in (3), we can determine that $k_0(t)+ k_1(t)= 0$; for $t\in\{0,9,20,21,29,30\}$, if we obtain a nonzero cube sum in (4), we can determine that $k_0(t)+ k_1(t)= 0$; for $t\in\{0,1,\ldots,63\}$ while $t\neq0,9,20,21,29,30,56,57,58,59$, if we obtain a nonzero cube sum in (5), we can determine that $k_0(t)+ k_1(t)= 0$; for $t\in\{56,57,58,59\}$, if we obtain a nonzero cube sum in (5), we can determine that $k_0(t)+ k_1(t)= 1$.

\begin{algorithm}
\caption{Tester in Attack on 5/6-round Initialization of \textsc{Ascon}. \\
\emph{Note: cube sum $C_i[t]$ means sum of the output of 5/6-round \textsc{Ascon} over cube variables in $(i)$ of Table~\ref{tab:5rattack}/~\ref{tab:6rattack}; $k_0$, $k_0+ k_1$ record values of key bits recovered; $flag[0][t]=1$ represents $k_0(t)$ has been recovered, $flag[0][t]=0$ represents $k_0(t)$ hasn't been recovered, and $flag[1][t]$ for $k_0(t)+ k_1(t)$ similarly; $remain$ represents the number of key bits that need to be searched exhaustively.}}
\label{alg:testor56}
\vspace{.1cm}
\begin{algorithmic}
\REQUIRE
$k_0,k_1$,
$flag[2][64]$ all set to 0
\ENSURE
$remain$, $k_0, k_0 + k_1$ and $flag[2][64]$\\
all elements of $k_0, k_0 + k_1,remain,flag[2][64]$ are set to 0
\FOR{$t\in\{0,1,...,63\}$}
    \STATE compute cube sum $C_1[t]$;
    \IF{$C_1[t]\not= 0$}
        \STATE $k_0(t)=1$;~$flag[0][t]=1$;
    \ELSE
        \STATE compute cube sum $C_2[t]$;
        \IF{$C_2[t]\not= 0$}
            \STATE $k_0(t)=0$;~$flag[0][t]=1$;
        \ENDIF
    \ENDIF
    \STATE compute cube sum $C_3[t]$;
    \IF{$C_3[t]\not= 0$}
        \IF{$t\in\{56,57,58,59\}$}
            \STATE $k_0(t)+ k_1(t)=0$;~$flag[1][t]=1$;
        \ELSE
            \STATE $k_0(t)+ k_1(t)=1$;~$flag[1][t]=1$;
        \ENDIF
    \ELSE
        \IF{$t\in \{0,9,20,21,29,30\}$}
            \STATE compute cube sum $C_4[t]$;
            \IF{$C_4[t]\not= 0$}
                \STATE $k_0(t)+ k_1(t)=0$;~$flag[1][t]=1$;
            \ENDIF
        \ELSE
            \STATE compute cube sum $C_5[t]$;
            \IF{$C_5[t]\not= 0$}
                \IF{$t\in\{56,57,58,59\}$}
                    \STATE $k_0(t)+ k_1(t)=1$;~$flag[1][t]=1$;
                \ELSE
                    \STATE $k_0(t)+ k_1(t)=0$;~$flag[1][t]=1$;
                \ENDIF
            \ENDIF
        \ENDIF
    \ENDIF
\ENDFOR

\FOR{$i\in\{0,1\}$}
    \FOR{$t\in\{0,1\ldots63\}$}
        \IF{$flag[i][t]==0$}
            \STATE $remain$++;
        \ENDIF
    \ENDFOR
\ENDFOR
\RETURN $remain, k_0,  k_0 + k_1$ and $flag[2][64]$;
\end{algorithmic}
\end{algorithm}

Computation of a cube sum need $2^{16}$ (nonce, $P_1\oplus C_1$) (as shown in Figure \ref{fig:ASCONpart}) pairs, and each pair corresponds to an operation of initialization of \textsc{Ascon}. Thus, time and data share the same complexity in cube computation. As each cube sum consumes $2^{16}$, tester described in Algorithm~\ref{alg:testor56} with Table~\ref{tab:5rattack} needs at most $64\times2\times2\times2^{16}=2^{24}$. In our experiments, $remain$ returned by Algorithm~\ref{alg:testor56} with Table~\ref{tab:5rattack} is always less than 14. For $2^{14}$ is far less than $2^{24}$, the whole time complexity is $2^{24}$. The data complexity is $2^{24}$ as well. Source code is in \url{https://github.com/lizhengcn/Ascon_test}.

\section{Attack on 6-round initialization of \textsc{Ascon}}\label{sec:6rASCON}
Similar to the attack on 5-round, we choose cube $\{v_{0},v_{1}... v_{31}\}$ obeying the following rules:

1. Any two of $\{v_{0},v_{1}... v_{31}\}$ do not multiply in the S-box operation of the first round.

2. If $k_0(t)$=0, $v_{0}$ doesn't multiply with any of $\{v_{1},v_{2}... v_{31}\}$ in the S-box operation of the second round.

3. If $k_0(t)$=1, $v_{0}$ multiplies with some of $\{v_{1},v_{2}... v_{31}\}$ in the S-box operation of the second round.

Thus, if $k_0(t)$=0, no $v_0v_i, i\in\{1,2,...,31\}$, exists in $S_2$, and $v_{0}v_{1}... v_{31}$ does not appear in $S_6$. However, if $k_0(t)=1$, $v_{0}v_{1}... v_{31}$ will appear in $S_6$ with a high probability. It means that $g=k_0(t)$ in Corollary \ref{cor:cubecorollary}. Therefore, if we obtain a nonzero cube sum of the output of 6-round \textsc{Ascon} over $v_{0},v_{1}... v_{31}$ with parameters in (1) of Table~\ref{tab:6rattack}, we can determine that $k_0(t)=1$, which is identical to Property \ref{pro:cubesumk0}.
The whole procedure to recover the whole key is similar to the attack on 5-round initialization of \textsc{Ascon} in Section \ref{sec:5rASCON} applying Algorithm \ref{alg:testor56} with parameters in Table \ref{tab:6rattack}.

As each cube sum consumes $2^{32}$ (time and data as shown in Section \ref{sec:5rASCON}), tester described in Algorithm~\ref{alg:testor56} with Table~\ref{tab:6rattack} needs at most $64\times2\times2\times2^{32}=2^{40}$. In our experiments, $remain$ returned by Algorithm~\ref{alg:testor56} with Table~\ref{tab:6rattack} is always 0. The whole time complexity is $2^{40}$. The data complexity is $2^{40}$ as well. Source code is in \url{https://github.com/lizhengcn/Ascon_test}.
\section{Attack on 7-round initialization of \textsc{Ascon}}\label{sec:7rASCON}

In this section, we apply the so-called \emph{the cube-like key-subset technique} to the 7-round key-recovery attack on \textsc{Ascon}. We find a series sets of \emph{partial divisors} related to the key bits for many 65-dimension cubes. That means, if the \emph{partial divisors} in a set equal to zero, the cube sums of the corresponding 65-dimension cube will be zero. The full key space is divided into many subsets indexed by the key bits in each set of \emph{partial divisors}. Then \emph{the cube-like key-subset technique} is applied to recover the key.

In detail, we set $S_{0}[3][j] = v_{j}$ for $j=0,1\ldots 63$ and $S_{0}[4][i] = v_{64}$ where $i$ could take a value from $\{0,1\ldots 63\}$. We denote the 65-dimension cube set as \emph{original cube set}. As $v_{0},v_{1},\ldots, v_{63}$ belong to distinct S-boxes, they do not multiply with each other in the first round. While $v_{64}$ and $v_{i}$ lie in the same $i$th S-box, $v_{i}v_{64}$ will be the only quadratic term after the first substitution layer. Based on Property~\ref{property:onlyx2}, $v_{i}v_{64}$ just appear in $S_{0.5}[2][i]$ after the S-box in the first round. The first linear diffusion layer makes the ANF of $S_{1}[2][i]$, $S_{1}[2][i+1]$\footnote{$i+1$ is short for $(i+1)mod~64$ , similarly $i+x$ is short for $(i+x) mod~64$ in this section.} and $S_{1}[2][i+6]$ contain $v_{i}v_{64}$. According to our \textbf{Test 1} and \textbf{Test 2} in Section~\ref{sec:rationality}, the monomial $v_0v_1\cdots v_{64}$ will appear in the output bits of 7-round. However, by assigning some key bit conditions, the quadratic terms $v_{i}v_{64}$ will not multiply with other cube variables in the second round. Thus, after the second round, only quadratic and linear terms exist (no terms exist with degree more than 2). Then in the output bits of 7-round, the degree of the product of cube variables is at most 64. In fact, the assigned key bit conditions lead all the  coefficients of the cubic terms $v_{i}v_{64}v_{j'}$ ($j'=0,1\ldots 63$ and $j'\neq i$) after the second round to be zero, which makes the disappearance of the monomial of degree 65 in the output bits of 7-round.
\textbf{Those coefficients are actually the so-called \emph{partial divisors} of the cube sums of the 65-dimension cube.}
In the following, we will extract the \emph{partial divisors} along with key bit conditions.

Based on Property~\ref{property:x2nox0x4}, only $S_{1}[1][i]$, $S_{1}[3][i]$ multiply with $S_{1}[2][i]$, and similarly
$S_{1}[1][i+1]$, $S_{1}[3][i+1]$ multiply with $S_{1}[2][i+1]$, and $S_{1}[1][i+6]$, $S_{1}[3][i+6]$ multiply with $S_{1}[2][i+6]$.
In detail, for the input of the $(i+1)$th S-box of second round, we list the ANF of $S_{1}[1][i+1]$, $S_{1}[3][i+1]$ and $S_{1}[2][i+1]$ in Eq.~(\ref{eq:original}) as an example, similar equations can be obtained for the input of the $i$th and $(i+6)$th S-box. Note that  $S_{0}[4][t] = n(t)$ for $t\in\{0,1\ldots 63\}$ while $t\neq i$.
According to Property~\ref{property:onlyx2} and~\ref{property:x2nox0x4}, in $(i+1)$th S-box of second round, only cubic terms $v_{i}v_{64}v_{i+1}$, $v_{i}v_{64}v_{i+4}$, $v_{i}v_{64}v_{i+26}$, $v_{i}v_{64}v_{i+48}$, $v_{i}v_{64}v_{i+55}$ are possibly produced by terms underlined in Eq.~(\ref{eq:original}). We display all the possible cubic terms in $S_{1.5}$ and their corresponding coefficients in Table~\ref{tab:original}.

\begin{eqnarray} \label{eq:original}
\begin{aligned}
&S_{1}[1][i+1]=\underline{(k_0(i+1)+k_1(i+1)+1)*v_{i+1} + (k_0(i+4)+k_1(i+4)+1)*v_{i+4}}\\
&\underline{ + (k_0(i+26)}\underline{+k_1(i+26)+1)*v_{i+26}} + \textbf{\emph{n(i+1) + n(i+4) + n(i+26)}} +  k_0(i+1)*\\
&k_1(i+1) + k_0(i+1) + k_0(i+4)*k_1(i+4) + k_0(i+4) + k_0(i+26)*k_1(i+26) +  \\
&k_0(i+26) + k_1(i+1) + k_1(i+4) + k_1(i+26) + IV(i+1) + IV(i+4) + IV(i+26)\\[2mm]
&S_{1}[2][i+1]=\underline{v_{i}*v_{64}} + k_0(i) + k_0(i+1) + k_0(i+59) + k_1(i) + k_1(i+1) + k_1(i+ \\
&59) + n(i+1)*v_{i+1} + n(i+1) + n(i+59)*v_{i+59} + n(i+59) + v_{i+64}\\[2mm]
&S_{1}[3][i+1]=\underline{(IV(i+1)+1)*v_{i+1} + (IV(i+48)+1)*v_{i+48} + (IV(i+55)+1)*}\\
&\underline{v_{i+55}} + \textbf{\emph{(IV(i+1)+1)*n({i+1}) + (IV(i+48)+1)*n({i+48}) + (IV(i+55)+1)*}}\\
&\textbf{\emph{n(i+55)}}+ k_0(i+1) + k_0(i+48) + k_0(i+55) + k_1(i+1) + k_1(i+48) + k_1(i+55) + \\
&IV(i+1)+ IV(i+48) + IV(i+55)
\end{aligned}
\end{eqnarray}

\begin{table}
\centering
\begin{tabular}{|c|c|c|}
\hline
\multirow{2}{*}{index of S-box} & \multirow{2}{*}{cubic terms}& corresponding coefficients\\
&&(\emph{partial divisors})\\
\hline
\multirow{6}{*}{$i+1$}&\multirow{2}{*}{$v_{i}v_{64}v_{i+1}$} & $k_0(i+1)+k_1(i+1)+1$\\
\cline{3-3}
&&$k_0(i+1)+k_1(i+1)+IV(i+1)$\\
\cline{2-3}
&$v_{i}v_{64}v_{i+4}$ & $k_0(i+4)+k_1(i+4)+1$\\
\cline{2-3}
&$v_{i}v_{64}v_{i+26}$& $k_0(i+26)+ k_1(i+26)+1$\\
\cline{2-3}
&$v_{i}v_{64}v_{i+48}$& $IV(i+48)+1$\\
\cline{2-3}
&$v_{i}v_{64}v_{i+55}$& $IV(i+55)+1$\\
\hline
\multirow{4}{*}{$i$}&$v_{i}v_{64}v_{i+3}$ & $k_0(i+3)+k_1(i+3)+1$\\
\cline{2-3}
&$v_{i}v_{64}v_{i+25}$ & $k_0(i+25)+k_1(i+25)+1$\\
\cline{2-3}
&$v_{i}v_{64}v_{i+47}$ & $IV(i+47)+1$\\
\cline{2-3}
&$v_{i}v_{64}v_{i+54}$ & $IV(i+54)+1$\\
\hline
\multirow{6}{*}{$i+6$}&\multirow{2}{*}{$v_{i}v_{64}v_{i+6}$} & $k_0(i+6)+k_1(i+6)+1$\\
\cline{3-3}
&&$k_0(i+6)+k_1(i+6)+IV(i+6)$\\
\cline{2-3}
&$v_{i}v_{64}v_{i+9}$ & $k_0(i+9)+k_1(i+9)+1$\\
\cline{2-3}
&$v_{i}v_{64}v_{i+31}$ & $k_0(i+31)+k_1(i+31)+1$\\
\cline{2-3}
&$v_{i}v_{64}v_{i+53}$ & $IV(i+53)+1$\\
\cline{2-3}
&$v_{i}v_{64}v_{i+61}$ & $IV(i+60)+1$\\
\hline
\noalign{\smallskip}
\end{tabular}
\caption{Coefficients of Cubic Terms with No Additional Cube Set}
\label{tab:original}
\end{table}
\pagebreak

\noindent\textbf{CASE 1:  all the \emph{IV} bits in Table~\ref{tab:original} are 0.}

As $IV(j)=0$ for $j\in\{0,1\ldots63\}$ while $j\neq0,9,20,21,29,30$, we firstly consider the case that all the relative bits in $IV$ are 0. To achieve a zero cube sum, all the coefficients in Table~\ref{tab:original} should be 0.
However, in Table \ref{tab:original}, if all the relative $IV$ bits are 0, then the coefficients $k_0(i+1)+k_1(i+1)+1$,  $k_0(i+1)+k_1(i+1)+IV(i+1)$ can not be zero simultaneously, and $IV(i+48)+1$, $IV(i+55)+1$ etc are certain to equal 1. To solve this problem, we introduce some auxiliary cube variables to zero the coefficients. These auxiliary cube variables are introduced as follows in Definition \ref{def:addcubeset}.
\begin{Definition}\label{def:addcubeset}
For any value of key, some additional nonce bits in $S_{0}[4]$ should be set as cube variables to eliminate the cases that no key can zero all the coefficients in Table
\ref{tab:original}, the additional cube variable is called auxiliary cube variable. Auxiliary cube variable is necessary to nullify cubic terms in $S_{1.5}$. It should be noted that $S_{0}[4][j]$ is set as $v_j$ which is equal to $S_{0}[3][j]$ for some given $j\in\{0,1,...,63\}$.
\end{Definition}
If the added auxiliary cube variables are $S_{0}[4][i+1]=n(i+1)=v_{i+1},S_{0}[4][i+48]=n(i+48)=v_{i+48},S_{0}[4][i+55]=n(i+55)=v_{i+55}$,
in the ANF of $S_{1}[3][i+1]$ of Eq.~(\ref{eq:original}), parts of the underline formula and the bold formula will be equal and cancelled out. Then the Eq.~(\ref{eq:original}) is simplified as follow in Eq.~(\ref{eq:addbasic}).
\begin{eqnarray} \label{eq:addbasic}
\begin{aligned}
&S_{1}[1][i+1]=\underline{(k_0(i+1)+k_1(i+1))*v_{i+1} + (k_0(i+4)+k_1(i+4)+1)*v_{i+4}}\\
&\underline{ + (k_0(i+26)}\underline{+k_1(i+26)+1)*v_{i+26}} + \textbf{\emph{n(i+4) + n(i+26)}} +  k_0(i+1)*k_1(i+1)\\
&+ k_0(i+1) + k_0(i+4)*k_1(i+4) + k_0(i+4) + k_0(i+26)*k_1(i+26) + k_0(i+26) \\
&+ k_1(i+1) + k_1(i+4) + k_1(i+26) + IV(i+1) + IV(i+4) + IV(i+26)\\[2mm]
&S_{1}[2][i+1]=\underline{v_{i}*v_{64}} + k_0(i) + k_0(i+1) + k_0(i+59) + k_1(i) + k_1(i+1) + \\
&k_1(i+59) + n(i+1)*v_{i+1} + n(i+1) + n(i+59)*v_{i+59} + n(i+59) + v_{i+64}\\[2mm]
&S_{1}[3][i+1]=k_0(i+1) + k_0(i+48) + k_0(i+55) + k_1(i+1) + k_1(i+48) + k_1(i+55)\\
&IV(i+1)+ IV(i+48) + IV(i+55)
\end{aligned}
\end{eqnarray}
Then, compared to Table~\ref{tab:original}, the coefficients of $v_{i}v_{64}v_{i+48}$ and $v_{i}v_{64}v_{i+55}$ turn to 0, and $k_0(i+1)+k_1(i+1)$ becomes the only coefficient of $v_{i}v_{64}v_{i+1}$ instead of $k_0(i+1)+k_1(i+1)+1$ and $k_0(i+1)+k_1(i+1)+IV(i+1)$.
Additionally, we can add more auxiliary cube variables to zero more coefficients as shown in the new coefficients Table~\ref{tab:addbasic}. Hence, the nonzero coefficients of the cubic terms are all determined by some key bits. We set these coefficients to be zero and get a equation set Eq.~(\ref{eq:basickey}).
Hence, if a secret key conforms to Eq.~(\ref{eq:basickey}), then all the coefficients in Table~\ref{tab:addbasic} are 0 and there will be no cubic terms after second round.
\begin{table}
\centering
\begin{tabular}{|c|c|c|c|}
\hline
\multirow{2}{*}{index of S-box} & \multirow{2}{*}{cubic terms} & \multirow{2}{*}{auxiliary cube variables} &corresponding coefficients\\
&&&(\emph{partial divisors})\\
\hline
\multirow{5}{*}{$i+1$}& $v_{i}v_{64}v_{i+1}$ & $S_{0}[4][i+1]=v_{i+1}$ & $k_0(i+1)+k_1(i+1)$\\
\cline{2-4}
&$v_{i}v_{64}v_{i+4}$ && $k_0(i+4)+k_1(i+4)+1$\\
\cline{2-4}
&$v_{i}v_{64}v_{i+26}$&& $k_0(i+26)+ k_1(i+26)+1$\\
\cline{2-4}
&$v_{i}v_{64}v_{i+48}$& $S_{0}[4][i+48]=v_{i+48}$ & $0$\\
\cline{2-4}
&$v_{i}v_{64}v_{i+55}$& $S_{0}[4][i+55]=v_{i+55}$ & $0$\\
\hline
\multirow{4}{*}{$i$}&$v_{i}v_{64}v_{i+3}$ && $k_0(i+3)+k_1(i+3)+1$\\
\cline{2-4}
&$v_{i}v_{64}v_{i+25}$ && $k_0(i+25)+k_1(i+25)+1$\\
\cline{2-4}
&$v_{i}v_{64}v_{i+47}$ & $S_{0}[4][i+47]=v_{i+47}$ & $0$\\
\cline{2-4}
&$v_{i}v_{64}v_{i+54}$ & $S_{0}[4][i+54]=v_{i+54}$ & $0$\\
\hline
\multirow{5}{*}{$i+6$}&$v_{i}v_{64}v_{i+6}$ & $S_{0}[4][i+6]=v_{i+6}$ & $k_0(i+6)+k_1(i+6)$\\
\cline{2-4}
&$v_{i}v_{64}v_{i+9}$ && $k_0(i+9)+k_1(i+9)+1$\\
\cline{2-4}
&$v_{i}v_{64}v_{i+31}$ && $k_0(i+31)+k_1(i+31)+1$\\
\cline{2-4}
&$v_{i}v_{64}v_{i+53}$ & $S_{0}[4][i+53]=v_{i+53}$ & $0$\\
\cline{2-4}
&$v_{i}v_{64}v_{i+61}$ & $S_{0}[4][i+60]=v_{i+60}$ & $0$\\
\hline
\noalign{\smallskip}
\end{tabular}
\caption{Coefficients of Cubic Terms with Auxiliary Cube Variables}
\label{tab:addbasic}
\end{table}

\begin{eqnarray}\label{eq:basickey}
\left\{
\begin{aligned}
&k_0(i+\ 1)+k_1(i+\ 1)=0\\
&k_0(i+\ 4)+k_1(i+\ 4)=1\\
&k_0(i+26)+k_1(i+26)=1\\
&k_0(i+\ 3)+k_1(i+\ 3)=1\\
&k_0(i+25)+k_1(i+25)=1\\
&k_0(i+\ 6)+k_1(i+\ 6)=0\\
&k_0(i+\ 9)+k_1(i+\ 9)=1\\
&k_0(i+31)+k_1(i+31)=1
\end{aligned}
\right.
\end{eqnarray}
In order to get more different key conditions that delete the cubic terms, we introduce the control cube variable which is defined as follows.
\begin{Definition}\label{def:addcontrolcubeset}
When some additional nonce bit in $S_{0}[4]$ set as cube variable, the value of coefficients in Table \ref{tab:addbasic} will be XORed by 1. The additional cube variable is called \emph{control cube variable}. Control cube variable is tweakable to provide different coefficients of cubic terms in $S_{1.5}$ according to the recovery of different keys. $S_{0}[4][j]$ is set as $v_j$ which is equal to $S_{0}[3][j]$ for some given $j\in\{0,1,...,63\}$.
\end{Definition}
For example, we add $S_{0}[4][i+4]= n(i+4)=v_{i+4}$ as a control cube variable, the Eq. ~(\ref{eq:addbasic}) is changed to be Eq. (\ref{eq:addcontrol}). In the ANF of $S_{1}[1][i+1]$ of Eq. (\ref{eq:addbasic}), the bold formula $\textbf{\emph{n(i+4)}}$ and $(k_0(i+4)+k_1(i+4)+1)*v_{i+4}$ are added to be $(k_0(i+4)+k_1(i+4))*v_{i+4}$. Obviously, the coefficient is changed.

\begin{eqnarray} \label{eq:addcontrol}
\begin{aligned}
&S_{1}[1][i+1]=\underline{(k_0(i+1)+k_1(i+1))*v_{i+1} + (k_0(i+4)+k_1(i+4))*v_{i+4}}\\
&\underline{ + (k_0(i+26)}\underline{+k_1(i+26)+1)*v_{i+26}} + \textbf{\emph{n(i+26)}} +  k_0(i+1)*k_1(i+1)\\
&+ k_0(i+1) + k_0(i+4)*k_1(i+4) + k_0(i+4) + k_0(i+26)*k_1(i+26) + k_0(i+26) \\
&+ k_1(i+1) + k_1(i+4) + k_1(i+26) + IV(i+1) + IV(i+4) + IV(i+26)\\[2mm]
&S_{1}[2][i+1]=\underline{v_{i}*v_{64}} + k_0(i) + k_0(i+1) + k_0(i+59) + k_1(i) + k_1(i+1) + \\
&k_1(i+59) + n(i+1)*v_{i+1} + n(i+1) + n(i+59)*v_{i+59} + n(i+59) + v_{i+64}\\[2mm]
&S_{1}[3][i+1]=k_0(i+1) + k_0(i+48) + k_0(i+55) + k_1(i+1) + k_1(i+48) + k_1(i+55)
\end{aligned}
\end{eqnarray}

\textbf{Differences among \emph{original cube variables}, \emph{auxiliary cube variables} and \emph{control cube variables}:} We firstly select a 65-dimension \emph{original cube} which occupies $S_0[3]$ and 1-bit of $S_0[4]$. After the second round, cubic terms appear with coefficients determined by key bits and $IV$ bits. We would like to zero those coefficients, however, some coefficients can not be zero, e.g. the coefficient of  $v_iv_{64}v_{i+48}$ is $IV(i+55)+1$ while $IV(i+55)=0$, shown in Table \ref{tab:original}. Hence, \emph{auxiliary cube variables} are introduced to zero these shown in Table \ref{tab:addbasic}. Then we get key condition equation set -- Eq. (\ref{eq:basickey}) to zero those coefficients. When the secret key meets Eq. (\ref{eq:basickey}), the cube sums will be zero. However, it is just a weak key attack that covers a key subset of size $2^{120}$. We want get many more such key subsets to extend the weak key attack to a full key attack. Hence, the tweakable \emph{control cube variables} are introduced to provide different coefficients of cubic terms and get different key condition equation sets to cover different key subsets. So when an \emph{original cube} is fixed, the \emph{auxiliary cube variables} are also fixed, but \emph{control cube variables} are tweakable to cover different key subsets. Those are the differences between the three cube variables.

\begin{table}
\centering
\begin{tabular}{|c|c|c|c|c|}
\hline
 & cubic & auxiliary cube & control cube  &corresponding\\
 &terms& variables&variable& coefficients\\
\hline
\multirow{5}{*}{$i+1$}& $v_{i}v_{64}v_{i+1}$ & $S_{0}[4][i+1]=v_{i+1}$ && $k_0(i+1)+k_1(i+1)$\\
\cline{2-5}
&$v_{i}v_{64}v_{i+4}$ && $S_{0}[4][i+4]=v_{i+4}$ &$k_0(i+4)+k_1(i+4)$\\
\cline{2-5}
&$v_{i}v_{64}v_{i+26}$&&& $k_0(i+26)+ k_1(i+26)+1$\\
\cline{2-5}
&$v_{i}v_{64}v_{i+48}$& $S_{0}[4][i+48]=v_{i+48}$ && $0$\\
\cline{2-5}
&$v_{i}v_{64}v_{i+55}$& $S_{0}[4][i+55]=v_{i+55}$ && $0$\\
\hline
\multirow{4}{*}{$i$}&$v_{i}v_{64}v_{i+3}$ &&& $k_0(i+3)+k_1(i+3)+1$\\
\cline{2-5}
&$v_{i}v_{64}v_{i+25}$ &&& $k_0(i+25)+k_1(i+25)+1$\\
\cline{2-5}
&$v_{i}v_{64}v_{i+47}$ & $S_{0}[4][i+47]=v_{i+47}$ && $0$\\
\cline{2-5}
&$v_{i}v_{64}v_{i+54}$ & $S_{0}[4][i+54]=v_{i+54}$ && $0$\\
\hline
\multirow{5}{*}{$i+6$}&$v_{i}v_{64}v_{i+6}$ & $S_{0}[4][i+6]=v_{i+6}$ && $k_0(i+6)+k_1(i+6)$\\
\cline{2-5}
&$v_{i}v_{64}v_{i+9}$ &&& $k_0(i+9)+k_1(i+9)+1$\\
\cline{2-5}
&$v_{i}v_{64}v_{i+31}$ &&& $k_0(i+31)+k_1(i+31)+1$\\
\cline{2-5}
&$v_{i}v_{64}v_{i+53}$ & $S_{0}[4][i+53]=v_{i+53}$ && $0$\\
\cline{2-5}
&$v_{i}v_{64}v_{i+61}$ & $S_{0}[4][i+60]=v_{i+60}$ && $0$\\
\hline
\noalign{\smallskip}
\end{tabular}
\caption{Coefficients of Cubic Terms with Auxiliary and Control Cube Variable}
\label{tab:addcontrol}
\end{table}
We list coefficients of cubic terms in Table~\ref{tab:addcontrol} after adding control cube variable $S_{0}[4][i+4]= n(i+4)=v_{i+4}$. Compared to Table~\ref{tab:addbasic}, only $k_0(i+4)+k_1(i+4)+1$ is changed to $k_0(i+4)+k_1(i+4)$.
Key conditions required for a zero cube sum are adjusted to Eq.~(\ref{eq:controlkey}) from Eq.~(\ref{eq:basickey}) by a control cube variable $S_{0}[4][i+4]=v_{i+4}$.
\begin{eqnarray}\label{eq:controlkey}
\left\{
\begin{aligned}
&k_0(i+\ 1)+k_1(i+\ 1)=0\\
&k_0(i+\ 4)+k_1(i+\ 4)=0\\
&k_0(i+26)+k_1(i+26)=1\\
&k_0(i+\ 3)+k_1(i+\ 3)=1\\
&k_0(i+25)+k_1(i+25)=1\\
&k_0(i+\ 6)+k_1(i+\ 6)=0\\
&k_0(i+\ 9)+k_1(i+\ 9)=1\\
&k_0(i+31)+k_1(i+31)=1
\end{aligned}
\right.
\end{eqnarray}

Similarly, for $t\in\{3,4,9,25,26,31\}$, control cube variable $S_{0}[4][i+t]=v_{i+t}$ can change the corresponding kind of coefficients from $k_0(i+t)+k_1(i+t)+1$ to $k_0(i+t)+k_1(i+t)$. Therefore, there are $2^{6}=64$ kinds of control cube variable combinations corresponding to 64 groups of coefficients respectively. Hence, there are 64 different equation sets correspondingly shown in Eq.~(\ref{eq:controlkeyall}), where $(a,b,c,d,e,f)\in F_2^{6}$ varies according to different control cube variable combination.

\begin{eqnarray}\label{eq:controlkeyall}
\left\{
\begin{aligned}
&k_0(i+\ 1)+k_1(i+\ 1)=0\\
&k_0(i+\ 4)+k_1(i+\ 4)=a\\
&k_0(i+26)+k_1(i+26)=b\\
&k_0(i+\ 3)+k_1(i+\ 3)=c\\
&k_0(i+25)+k_1(i+25)=d\\
&k_0(i+\ 6)+k_1(i+\ 6)=0\\
&k_0(i+\ 9)+k_1(i+\ 9)=e\\
&k_0(i+31)+k_1(i+31)=f
\end{aligned}
\right.
\end{eqnarray}
When key meets the corresponding conditions, all the coefficients of cubic terms will be zero. It means that there are no cubic terms in $S_{1.5}$. Due to linearity of the linear diffusion layer, no cubic terms exist in $S_{2}$. The highest degree of monomials in $S_{2}$ is 2, which has also been verified using SAGE \cite{stein2005sage} by computer. As the algebraic degree of S-box is 2, degree of a term in the output of 7-round \textsc{Ascon} is at most $2^{5}=32$ times of the highest degree in $S_{2}$ (i.e. 2). Thus, the algebraic degree of the 7-round \textsc{Ascon}'s output is less than or equal to 64, which means that $v_{0}v_{1}\ldots v_{64}$ will not appear in the output, then the cube sum of cube $\{v_{0},v_{1},\ldots, v_{64}\}$ is zero. When key doesn't meet the corresponding conditions, some cubic terms will appear in $S_{2}$. Therefore, $v_{0}v_{1}\ldots v_{64}$ will appear in the output of 7-round.

In details,
for a key subset where the key meets the two equations of $k_0(i+1)+k_1(i+1)=0$ and $k_0(i+6)+k_1(i+6)=0$ in Eq.~(\ref{eq:controlkeyall}), we determine the values of $(a,b,c,d,e,f)$ by testing the cube sum of each of $64$ 65-dimension cubes indexed by control cube variable combination. The time complexity of the test is $2^{65}\times 64=2^{71}$. We denote this key subset as $U_i$ which is determined by the two key conditions $k_0(i+1)+k_1(i+1)=0$ and $k_0(i+6)+k_1(i+6)=0$, whose size is $2^{128}\times \frac{1}{4}$. Else, we can know that $k_0(i+1)+k_1(i+1)$ and $k_0(i+6)+k_1(i+6)$ do not equal to 0 at the the same time. Note that the cube variable $v_{64}$ is placed in $S_0[4][i]$, where $i\in \{0,1,...,63\}$, however, in \textbf{CASE 1}, we suppose the related bits in $IV$ are zero, so there are 52 such $U_i$s in total that are determined by 52 sets of two bits key conditions.

\noindent\textbf{CASE 2:  some \emph{IV} bits in Table~\ref{tab:original} are 1.}

As shown in Table \ref{tab:original}, if $IV(i+1)=1$, then the coefficients  $k_0(i+1)+k_1(i+1)+1$ and $k_0(i+1)+k_1(i+1)+IV(i+1)$ are translated to the same one $k_0(i+1)+k_1(i+1)+1$. Then, there is no need to add the auxiliary cube variables $S_{0}[4][i+1]=v_{i+1}$. The freed auxiliary cube  variable can be used as control cube variable. In fact, if we use $S_{0}[4][i+1]=v_{i+1}$ as a control cube variable, $k_0(i+1)+k_1(i+1)+1$ becomes another coefficient that can be changed to $k_0(i+1)+k_1(i+1)$. Then Eq.~(\ref{eq:controlkeyall}) is translated to Eq.~(\ref{eq:controlkeyall0}), where $(a,b,c,d,e,f,g)\in F_2^{7}$ varies according to different control cube variable combinations.
Therefore, similar to the previous analysis, if a key  meets the one bit key condition of $k_0(i+6)+k_1(i+6)=0$, we can obtain the value of $(a,b,c,d,e,f,g)$ by testing the cube sum of the $2^{7}$ 65-dimension cubes.
We denote those keys as a key subset $U'_i$, whose size is $2^{127}$. Else, we claim $k_0(i+6)+k_1(i+6)=1$. The time complexity of the test is $2^{65}\times 2^{7}=2^{72}$.

\begin{eqnarray}\label{eq:controlkeyall0}
\left\{
\begin{aligned}
&k_0(i+\ 1)+k_1(i+\ 1)=a\\
&k_0(i+\ 4)+k_1(i+\ 4)=b\\
&k_0(i+26)+k_1(i+26)=c\\
&k_0(i+\ 3)+k_1(i+\ 3)=d\\
&k_0(i+25)+k_1(i+25)=e\\
&k_0(i+\ 6)+k_1(i+\ 6)=0\\
&k_0(i+\ 9)+k_1(i+\ 9)=f\\
&k_0(i+31)+k_1(i+31)=g
\end{aligned}
\right.
\end{eqnarray}

Since there are 6 bits of $IV$ are 1, i.e. $IV(i+1)=1$ where $(i+1)\in \{0,9,20,21,29,30\}$, there are 6 similar equation sets as Eq.~(\ref{eq:controlkeyall0}). For each equation set, there is a bit key condition correspondingly, which are listed as follows in Eq.~ (\ref{eq:controlkeyall01}) to (\ref{eq:controlkeyall06}):
\begin{align}\label{eq:controlkeyall01}
k_0(5)+k_1(5)=0\\\label{eq:controlkeyall02}
k_0(14)+k_1(14)=0\\\label{eq:controlkeyall03}
k_0(25)+k_1(25)=0\\\label{eq:controlkeyall04}
k_0(26)+k_1(26)=0\\\label{eq:controlkeyall05}
k_0(34)+k_1(34)=0\\\label{eq:controlkeyall06}
k_0(35)+k_1(35)=0
\end{align}
Hence, there are 6 such $U'_i$ key subsets corresponding to the above 6 one-bit key conditions, respectively.

Similarly, as shown in Table~\ref{tab:original}, if $IV(i+6)=1$, then the auxiliary cube variable $S_{0}[4][i+6]=v_{i+6}$ is translated as control cube variable. Then, Eq.~(\ref{eq:controlkeyall}) is translated as Eq.~(\ref{eq:controlkeyall1}), where $(a,b,c,d,e,f,g)\in F_2^{7}$ varies according to different control cube variable combination. We can not translate $S_{0}[4][i+1]=v_{i+1}$ and $S_{0}[4][i+6]=v_{i+6}$ to control cube variable simultaneously, because $IV(i+1)=1$ and $IV(i+6)=1$ can not hold simultaneously according to the given $IV$.
\begin{eqnarray}\label{eq:controlkeyall1}
\left\{
\begin{aligned}
&k_0(i+\ 1)+k_1(i+\ 1)=0\\
&k_0(i+\ 4)+k_1(i+\ 4)=a\\
&k_0(i+26)+k_1(i+26)=b\\
&k_0(i+\ 3)+k_1(i+\ 3)=c\\
&k_0(i+25)+k_1(i+25)=d\\
&k_0(i+\ 6)+k_1(i+\ 6)=e\\
&k_0(i+\ 9)+k_1(i+\ 9)=f\\
&k_0(i+31)+k_1(i+31)=g
\end{aligned}
\right.
\end{eqnarray}

Similarly, for different $IV(i+6)=1$ where $(i+6)\in \{0,9,20,21,29,30\}$, there are totally 6 similar equation sets as Eq.~(\ref{eq:controlkeyall1}). For each equation set, there is a bit key condition correspondingly, which are also listed as follows in Eq.~(\ref{eq:controlkeyall11}) to (\ref{eq:controlkeyall16}):

\begin{align}\label{eq:controlkeyall11}
  k_0(59)+k_1(59)=0\\\label{eq:controlkeyall12}
  k_0(4)+k_1(4)=0\\\label{eq:controlkeyall13}
  k_0(15)+k_1(15)=0\\\label{eq:controlkeyall14}
  k_0(16)+k_1(16)=0\\\label{eq:controlkeyall15}
  k_0(24)+k_1(24)=0\\\label{eq:controlkeyall16}
  k_0(25)+k_1(25)=0
\end{align}

Note that  Eq.~(\ref{eq:controlkeyall03}) and (\ref{eq:controlkeyall16}) are equal. So there are totally 12-1=11 such $U'_i$  key subsets, where $i\in \{ 63, 8, 19, 20, 28, 29, 58, 3,14,15,23\}$\footnote{The indexes $i$ represent the positions of cube variable $v_{64}$, i.e. $S_0[4][i] = v_{64}$}, corresponding to the above 11 different one-bit key conditions. Each $U'_i$ key subset's size is $2^{127}$.
In \textbf{CASE 1},  the indexes $i$ of the 52 $U_i$ key subsets are in $\{0,1,2,...,63\}-\{ 63, 8, 19, 20, 28, 29, 58, 3,14,15,23, 30\}$ .

The total attack procedures are as follows:
\begin{enumerate}[(i)]
  \item Test the cube sum of $2^{7}$ 65-dimension cubes obtained in CASE 2 and determine 8-bit key information. If all the cube sums are not zero, the key is not in the corresponding key subset $U'_i$.
  \item Repeat step (i) for 11 times according to different one bit key condition  shown in Eq.~(\ref{eq:controlkeyall01}) to Eq.~(\ref{eq:controlkeyall15}), which cover all $U'_i$s .
  \item Test the cube sum of $2^{6}$ 64-dimension cubes obtained in CASE 1  and determine 8-bit key information. If all the cube sums are not zero, then the key is not in the corresponding key subset $U_i$.
  \item Repeat step (iii) for 52 times to cover all $U_i$s.
\end{enumerate}

The time complexity of the above 4 procedures is $2^{7}\times 2^{65}\times 11 + 2^{6}\times 2^{65}\times 52 = 2^{77.2}$.

\textbf{In the worst case,} if the secret key does not fall into any $U_i$ and $U'_i$, then the key falls into a \emph{remaining key subset} whose size is about $2^{103.92}$. We explain the range of the \emph{remaining key subset} in details.

We propose a fast key filter phase to get the \emph{remaining key subset} in the following. We introduce a array of index $KI_0=\{5,14,25,26,34,35,59,4,15,16,24\}$. In the \emph{remaining key subset}, any key does not fall into any $U'_i$, which means any key does not obey Eq.~(\ref{eq:controlkeyall01}) to (\ref{eq:controlkeyall06}) and Eq.~(\ref{eq:controlkeyall11}) to (\ref{eq:controlkeyall16}). Hence, the keys meet the following conditions in Eq. (\ref{eq:controlkeydis1}):
\begin{equation}\label{eq:controlkeydis1}
  k_0(i)+k_1(i)=1, i\in KI_0.
\end{equation}

For each of $( k_0(i), k_1(i) ) (i\in KI_0)$, only 2 guesses out of $2^2$ values of $(k_0(i), k_1(i))$ obey the corresponding one of Eq.~(\ref{eq:controlkeydis1}). Each filtration ratio is $\frac{1}{2}$, so 11 $(k_0(i), k_1(i))$ lead a ratio $2^{-11}$. Additionally, any key in the \emph{remaining key subset} does not fall into any $U_i$, which means $k_0(i+1)+k_1(i+1)=0$ and $k_0(i+6)+k_1(i+6)=0$ can not be obeyed at the same time in Eq. (\ref{eq:controlkeyall}). It can be converted to the following tester:
\begin{equation}\label{eq.3/4tester}
 (k_0(i+1)+k_1(i+1)+1)(k_0(i+6)+k_1(i+6)+1)=0
\end{equation}
If we consider the four bits independently, $2^{4}$ key guesses over $k_0(i+1), k_1(i+1), k_0(i+6), k_1(i+6)$ will be filtered by Eq. (\ref{eq.3/4tester}) with filtration ratio of $\frac{3}{4}$.

As the index varies, it seems that 52 testers for key bits can be performed independently. However, the relevant key bits intersect for 52 testers. Firstly, we classify the indexes $i+1$ and $i+6$ for $k_0$ and $k_1$ in
Eq. (\ref{eq.3/4tester}) into the following arrays $KI_1$, $KI_2$ and $KI_3$.

\begin{eqnarray}\label{eq.ki1-ki3}
\left\{
\begin{aligned}
KI_{1}=\{&40, 45, 50, 55, 60, 1, 6, 11\}\\
KI_{2}=\{&31, 36, 41, 46, 51, 56, 61, 2, 7, 12, 17, 22, 27, 32, 37, 42,\\
 &47, 52, 57, 62, 3, 8, 13, 18,23, 28, 33, 38, 43, 48, 53, 58, 63\}\\
KI_{3}=\{&39, 44, 49, 54\}
\end{aligned}
\right.
\end{eqnarray}
$KI_j$ doesn't intersect with each other for $j\in\{0,1,2,3\}$. In array $KI_1$, suppose that $i_0$ is an arbitrary element and $i_1$ is next to it, then Eq. (\ref{eq.i0i1}) should be hold.
\begin{equation}\label{eq.i0i1}
 (k_0(i_0)+k_1(i_0)+1)(k_0(i_1)+k_1(i_1)+1)=0
\end{equation}
For $2^8$ guesses of $k_0(i)+k_1(i),i\in KI_1$, 55 values meet above conditions Eq.  (\ref{eq.i0i1}). So $KI_1$ provides a filtration ratio of $\frac{55}{2^8}$. Similarly, for $KI_j(j\in\{2,3\})$, the filtration ratios are $\frac{9227465}{2^{33}}, \frac{8}{2^4}$, respectively. Due to the disappearances of key in any $U_i$ of \textbf{CASE 1}, it provides a filtration ratio equals $\frac{55}{2^8}\times\frac{9227465}{2^{33}}\times \frac{8}{2^4}= 2^{-13.08}$.
The disappearance of key in any $U_i$ and $U'_i$ gives a total filtration ratio of $2^{-11}\times2^{-13.08}$. Since, the filter phase can be performed for each parts of the key independently, the time complexity is about $2^{11}$ checks of Eq. (\ref{eq:controlkeydis1}) and  $2^{8}+2^{33}+2^{4} = 2^{33.1}$ checks of Eq.  (\ref{eq.i0i1}).

After the filter phase, all the remaining keys should be checked by a $(nonce,plaintext,\\ciphertext,tag)$ pair to determine the right one.
The size \emph{remaining key subset} size is about $2^{128-11}\times 2^{-13.08}=2^{103.92}$. So in the worst case, the total complexity is $2^{11}+2^{33.1}+2^{103.92}+2^{77.2}=2^{103.92}$.

\textbf{In the best case,} if the secret key falls into the  intersection of all 11 $U'_i$s, denoted as \emph{weak-key set}, whose size is $2^{128-11}=2^{117}$.
52 bits\footnote{
Note that for  11 $U'_i$s, we get 11 sets of equations, such as Eq. (\ref{eq:controlkeyall1}). totally $11\times 8 = 88$ linear equations on key bits are retrieved. However, the linear equations may repeat, and only 52 linear equations are left.} key information are recovered and other  $128-52=76$ bits key are recovered by exhaustive search. The total time complexity to recover the right key for the $2^{117}$ \emph{weak-key set} is about $2^{76}+2^{7}\times 2^{65}\times 11 = 2^{77}$.
\section{Discussion on \textsc{Ascon}-128a and \textsc{Ascon} v1.1}\label{sec:ascon128a}
For attacks on 5/6-round initialization of \textsc{Ascon}-128a, with the same cubes and conditions to \textsc{Ascon}-128, key recovery can be performed in the same time and data complexity.

In the attack on 7-round initialization of \textsc{Ascon}, as \textbf{CASE 2} described in Section~\ref{sec:7rASCON}, it is considered specially when $IV$ bits are 1.
Since the $IV$s used by \textsc{Ascon}-128 and \textsc{Ascon}-128a are slightly different, the attack on 7-round \textsc{Ascon}-128a is a little different from that on \textsc{Ascon}-128. The details are described in Appendix \ref{app:ascon128a}. The total time complexity of the 7-round attack on \textsc{Ascon}-128a is about $2^{103.45}$.

\textsc{Ascon} v1.1 is a previous version. As the tweak only affects the ordering of constants in $p^b$, our attacks can also be applied to it.

\section{Conclusion}\label{sec:conclusion}
This paper improves the previous cube-like attack on \textsc{Ascon} by the \emph{generalized conditional cube attack} and \emph{the cube-like key-subset technique}. Different from previous cube-like method or conditional cube method \cite{DBLP:conf/eurocrypt/DinurMPSS15,cryptoeprint:2016:790,DBLP:conf/ctrsa/DobraunigEMS15}, we release the restriction that all cube variables must not multiply with each other in the first round. This makes it possible  for the 7-round key-recovery attack on \textsc{Ascon} which needs more cube variables to construct some new cubes with bigger dimension. In each of the new cubes, two cube variables multiply with each other to generate quadratic terms after the first round. In the second round, by restraining some bit conditions of the key and adding some auxiliary variables, the quadratic terms of cube variables do not multiply with other monomials of cube variables.
Interestingly, we finally divide the full key space into 63 key subsets and one remaining set according to different key bit conditions. In each subset, some 65-dimension cubes are constructed. Then, by testing the cube sum of these cubes, we determine which subsets the secret key falls into. If all the cube tests fail, we claim that the key falls into the remaining set and search it for the right one. This leads to a 7-round key-recovery attack on \textsc{Ascon} with time complexity $2^{103.92}$. Moreover, if the key falls into a weak-key set, whose size is $2^{117}$, the total complexity is reduced to $2^{77}$. We also give the first practical key-recovery attack on 6-round \textsc{Ascon}. Those are the best attacks on the round-reduced \textsc{Ascon}.

\section*{Acknowledgments}
We would like to thank Itai Dinur, Willi Meier, Senyang Huang and the anonymous reviewers who helped improve this paper. This work is supported by China's 973 Program (No. 2013CB834205), the Strategic Priority Research Program of the Chinese Academy of Sciences (No. XDB01010600), the National Natural Science Foundation of China (No. 61672019 and 61402256), the Fundamental Research Funds of Shandong University (No. 2016JC029), and the Foundation of Science and Technology on Information Assurance Laboratory (No. KJ-15-002).

\bibliography{reference}
\bibliographystyle{alpha}

\appendix
\renewcommand{\thesection}{\Alph{section}}
\renewcommand{\thesubsection}{\Alph{section}.\arabic{subsection}}

\section{Parameters sets}\label{app:parameters56}

Parameters sets for attacks on 5/6-round initialization of \textsc{Ascon} are listed in Table~\ref{tab:5rattack}/\ref{tab:6rattack} respectively.

\begin{table}
\caption{Parameters Set for Attack on the 5-round Initialization of \textsc{Ascon}}
\label{tab:5rattack}
\centering
\begin{tabular}{|c|c|l|}
\hline
\multicolumn{3}{|c|}{$k_0(t)$}\\
\hline
\multirow{10}{*}{(1)}& \multirow{5}{*}{cube variables}
& $S_0$[4][t]=$v_{0}$,$S_0$[4][5+t]$^{\dag}$=$v_{1}$,$S_0$[4][8+t]=$v_{2}$,$S_0$[4][12+t]=$v_{3}$,\\
&& $S_0$[4][14+t]=$v_{4}$,$S_0$[4][15+t]=$v_{5}$,$S_0$[4][18+t]=$v_{6}$,\\
&& $S_0$[4][19+t]=$v_{7}$,$S_0$[4][21+t]=$v_{8}$,$S_0$[4][27+t]=$v_{9}$,\\
&& $S_0$[4][28+t]=$v_{10}$,$S_0$[4][30+t]=$v_{11}$,$S_0$[4][34+t]=$v_{12}$,\\
&& $S_0$[4][37+t]=$v_{13}$,$S_0$[4][49+t]=$v_{14}$,$S_0$[4][50+t]=$v_{15}$.\\
\cline{2-3}
&\multirow{4}{*}{nonce}
&  $S_0$[3][t]=0,$S_0$[3][5+t]=0,$S_0$[3][8+t]=0,$S_0$[3][12+t]=0,\\
&& $S_0$[3][14+t]=0,$S_0$[3][15+t]=0,$S_0$[3][18+t]=0,$S_0$[3][19+t]=0,\\
&& $S_0$[3][21+t]=0,$S_0$[3][27+t]=0,$S_0$[3][28+t]=0,$S_0$[3][30+t]=0,\\
&& $S_0$[3][34+t]=0,$S_0$[3][37+t]=0,$S_0$[3][49+t]=0,$S_0$[3][50+t]=0.\\
\cline{2-3}
& key information & If a cube sum is nonzero, $k_0(t)= 1$.\\
\cline{1-3}
\multirow{10}{*}{(2)}& \multirow{5}{*}{cube variables}
&  $S_0$[4][t]=$v_{0}$,$S_0$[4][5+t]=$v_{1}$,$S_0$[4][7+t]=$v_{2}$,$S_0$[4][8+t]=$v_{3}$,\\
&& $S_0$[4][14+t]=$v_{4}$,$S_0$[4][15+t]=$v_{5}$,$S_0$[4][24+t]=$v_{6}$,\\
&& $S_0$[4][27+t]=$v_{7}$,$S_0$[4][30+t]=$v_{8}$,$S_0$[4][34+t]=$v_{9}$,\\
&& $S_0$[4][37+t]=$v_{10}$,$S_0$[4][41+t]=$v_{11}$,$S_0$[4][43+t]=$v_{12}$,\\
&& $S_0$[4][49+t]=$v_{13}$,$S_0$[4][50+t]=$v_{14}$,$S_0$[4][52+t]=$v_{15}$.\\
\cline{2-3}
&\multirow{4}{*}{nonce}
&  $S_0$[3][t]=0,$S_0$[3][5+t]=0,$S_0$[3][7+t]=0,$S_0$[3][8+t]=0,\\
&& $S_0$[3][14+t]=0,$S_0$[3][15+t]=0,$S_0$[3][24+t]=0,$S_0$[3][27+t]=0,\\
&& $S_0$[3][30+t]=0,$S_0$[3][34+t]=0,$S_0$[3][37+t]=0,$S_0$[3][41+t]=0,\\
&& $S_0$[3][43+t]=0,$S_0$[3][49+t]=0,$S_0$[3][50+t]=0,$S_0$[3][52+t]=0.\\
\cline{2-3}
& key information& If a cube sum is nonzero, $k_0(t)= 0$.\\
\hline\hline
\multicolumn{3}{|c|}{$k_0(t)+ k_1(t)^{\ddag}$}\\
\hline
\multirow{10}{*}{(3)}& \multirow{5}{*}{cube variables}
&  $S_0$[3][t]=$S_0$[4][t]=$v_{0}$,$S_0$[3][1+t]=$v_{1}$,$S_0$[3][4+t]=$v_{2}$,\\
&& $S_0$[3][5+t]=$v_{3}$,$S_0$[3][6+t]=$v_{4}$,$S_0$[3][8+t]=$v_{5}$,$S_0$[3][14+t]=$v_{6}$,\\
&& $S_0$[3][15+t]=$v_{7}$,$S_0$[3][16+t]=$v_{8}$,$S_0$[3][17+t]=$v_{9}$,\\
&& $S_0$[3][20+t]=$v_{10}$,$S_0$[3][26+t]=$v_{11}$,$S_0$[3][27+t]=$v_{12}$,\\
&& $S_0$[3][29+t]=$v_{13}$,$S_0$[3][30+t]=$v_{14}$,$S_0$[3][33+t]=$v_{15}$.\\
\cline{2-3}
&\multirow{4}{*}{nonce}
&  $S_0$[4][1+t]=$0$,$S_0$[4][4+t]=$0$,$S_0$[4][5+t]=$0$,$S_0$[4][6+t]=$0$,\\
&& $S_0$[4][8+t]=$0$,$S_0$[4][14+t]=$0$,$S_0$[4][15+t]=$0$,$S_0$[4][16+t]=$0$,\\
&& $S_0$[4][17+t]=$0$,$S_0$[4][20+t]=$0$,$S_0$[4][26+t]=$0$,$S_0$[4][27+t]=$0$,\\
&& $S_0$[4][29+t]=$0$,$S_0$[4][30+t]=$0$,$S_0$[4][33+t]=$0$.\\
\cline{2-3}
& key information & If a cube sum is nonzero, $k_0(t)+ k_1(t)= 1$.\\
\cline{1-3}
\multirow{6}{*}{(4)}& \multirow{5}{*}{cube variables}
&$S_0$[3][t]=$v_{0}$,$S_0$[3][5+t]=$v_{1}$,$S_0$[3][8+t]=$v_{2}$,$S_0$[3][14+t]=$v_{3}$,\\
&& $S_0$[3][15+t]=$v_{4}$,$S_0$[3][16+t]=$v_{5}$,$S_0$[3][17+t]=$v_{6}$,\\
&& $S_0$[3][20+t]=$v_{7}$,$S_0$[3][27+t]=$v_{8}$,$S_0$[3][29+t]=$v_{9}$,\\
&& $S_0$[3][30+t]=$v_{10}$,$S_0$[3][33+t]=$v_{11}$,$S_0$[3][34+t]=$v_{12}$,\\
&& $S_0$[3][35+t]=$v_{13}$,$S_0$[3][37+t]=$v_{14}$,$S_0$[3][38+t]=$v_{15}$.\\
\cline{2-3}
& key information& If a cube sum is nonzero, $k_0(t)+k_1(t)= 0$.\\
\hline
\multirow{6}{*}{(5)}& \multirow{5}{*}{cube variables}
&$S_0$[3][t]=$v_{0}$,$S_0$[3][5+t]=$v_{1}$,$S_0$[3][8+t]=$v_{2}$,$S_0$[3][14+t]=$v_{3}$,\\
&& $S_0$[3][15+t]=$v_{4}$,$S_0$[3][27+t]=$v_{5}$,$S_0$[3][29+t]=$v_{6}$,\\
&& $S_0$[3][30+t]=$v_{7}$,$S_0$[3][34+t]=$v_{8}$,$S_0$[3][36+t]=$v_{9}$,\\
&& $S_0$[3][37+t]=$v_{10}$,$S_0$[3][38+t]=$v_{11}$,$S_0$[3][39+t]=$v_{12}$,\\
&& $S_0$[3][45+t]=$v_{13}$,$S_0$[3][49+t]=$v_{14}$,$S_0$[3][50+t]=$v_{15}$.\\
\cline{2-3}
& key information& If a cube sum is nonzero, $k_0(t)+ k_1(t) = 0$ .\\
\hline
\multicolumn{3}{l}{$\dag$: (5+t) means (5+t) mod 64, similarly (x+t) means (x+t) mod 64 in Table \ref{tab:5rattack} and \ref{tab:6rattack}.}\\
\multicolumn{3}{l}{$\ddag$: $k_0(t)+ k_1(t)$ should reverse when $t\in\{56,57,58,59\}$ in Table \ref{tab:5rattack} and \ref{tab:6rattack}.}

\end{tabular}
\end{table}

\begin{table}
\centering
\begin{tabular}{|c|c|l|}
\hline
\multicolumn{3}{|c|}{$k_0(t)$}\\
\hline
\multirow{10}{*}{(1)}&
&$S_0$[3][t]=$S_0$[4][t]=$v_{0}$,$S_0$[3][1+t]=$v_{1}$,$S_0$[3][4+t]=$v_{2}$,$S_0$[3][5+t]=$v_{3}$,\\
&& $S_0$[3][6+t]=$v_{4}$,$S_0$[3][7+t]=$v_{5}$,$S_0$[3][8+t]=$v_{6}$,$S_0$[3][10+t]=$v_{7}$,\\
&& $S_0$[3][13+t]=$v_{8}$,$S_0$[3][14+t]=$v_{9}$,$S_0$[3][15+t]=$v_{10}$,$S_0$[3][16+t]=$v_{11}$,\\
&& $S_0$[3][17+t]=$v_{12}$,$S_0$[3][24+t]=$v_{13}$,$S_0$[3][26+t]=$v_{14}$,$S_0$[3][27+t]=$v_{15}$,\\
&cube& $S_0$[3][30+t]=$v_{16}$,$S_0$[3][34+t]=$v_{17}$,$S_0$[3][35+t]=$v_{18}$,$S_0$[3][37+t]=$v_{19}$,\\
&& $S_0$[3][40+t]=$v_{20}$,$S_0$[3][41+t]=$v_{21}$,$S_0$[3][43+t]=$v_{22}$,$S_0$[3][46+t]=$v_{23}$,\\
&& $S_0$[3][48+t]=$v_{24}$,$S_0$[3][49+t]=$v_{25}$,$S_0$[3][50+t]=$v_{26}$,$S_0$[3][52+t]=$v_{27}$,\\
&& $S_0$[3][56+t]=$v_{28}$,$S_0$[3][59+t]=$v_{29}$,$S_0$[3][60+t]=$S_0$[4][60+t]=$v_{30}$,\\
&& $S_0$[3][63+t]=$S_0$[4][63+t]=$v_{31}$.\\
\cline{2-3}
& key
& If a cube sum is nonzero, $k_0(t)= 1$.\\
\cline{1-3}
 \multirow{10}{*}{(2)}&
&  $S_0$[3][t]=$S_0$[4][t]=$v_{0}$,$S_0$[3][1+t]=$v_{1}$,$S_0$[3][4+t]=$v_{2}$,$S_0$[3][5+t]=$v_{3}$,\\
&& $S_0$[3][6+t]=$v_{4}$,$S_0$[3][8+t]=$v_{5}$,$S_0$[3][9+t]=$v_{6}$,$S_0$[3][12+t]=$v_{7}$,\\
&& $S_0$[3][14+t]=$v_{8}$,$S_0$[3][15+t]=$v_{9}$,$S_0$[3][16+t]=$v_{10}$,$S_0$[3][17+t]=$v_{11}$,\\
&& $S_0$[3][18+t]=$v_{12}$,$S_0$[3][19+t]=$v_{13}$,$S_0$[3][21+t]=$v_{14}$,$S_0$[3][26+t]=$v_{15}$,\\
&cube& $S_0$[3][27+t]=$v_{16}$,$S_0$[3][28+t]=$v_{17}$,$S_0$[3][30+t]=$v_{18}$,$S_0$[3][34+t]=$v_{19}$,\\
&& $S_0$[3][35+t]=$v_{20}$,$S_0$[3][37+t]=$v_{21}$,$S_0$[3][40+t]=$v_{22}$,$S_0$[3][46+t]=$v_{23}$,\\
&& $S_0$[3][48+t]=$v_{24}$,$S_0$[3][49+t]=$v_{25}$,$S_0$[3][50+t]=$v_{26}$,$S_0$[3][53+t]=$v_{27}$,\\
&& $S_0$[3][56+t]=$v_{28}$,$S_0$[3][59+t]=$v_{29}$,$S_0$[3][60+t]=$S_0$[4][60+t]=$v_{30}$,\\
&& $S_0$[3][63+t]=$S_0$[4][63+t]=$v_{31}$.\\
\cline{2-3}
&key
& If a cube sum is nonzero, $k_0(t)=0$.\\
\hline\hline
\multicolumn{3}{|c|}{$k_0(t)+ k_1(t)$}\\
\hline
\multirow{9}{*}{(3)}&
&  $S_0$[3][t]=$S_0$[4][t]=$v_{0}$,$S_0$[3][1+t]=$v_{1}$,$S_0$[3][4+t]=$v_{2}$,$S_0$[3][5+t]=$v_{3}$,\\
&& $S_0$[3][6+t]=$v_{4}$,$S_0$[3][8+t]=$v_{5}$,$S_0$[3][14+t]=$v_{6}$,$S_0$[3][15+t]=$v_{7}$,\\
&& $S_0$[3][16+t]=$v_{8}$,$S_0$[3][17+t]=$v_{9}$,$S_0$[3][20+t]=$v_{10}$,$S_0$[3][26+t]=$v_{11}$,\\
&& $S_0$[3][27+t]=$v_{12}$,$S_0$[3][29+t]=$v_{13}$,$S_0$[3][30+t]=$v_{14}$,$S_0$[3][33+t]=$v_{15}$,\\
&cube& $S_0$[3][34+t]=$v_{16}$,$S_0$[3][35+t]=$v_{17}$,$S_0$[3][37+t]=$v_{18}$,$S_0$[3][38+t]=$v_{19}$,\\
&& $S_0$[3][39+t]=$v_{20}$,$S_0$[3][40+t]=$v_{21}$,$S_0$[3][46+t]=$v_{22}$,$S_0$[3][48+t]=$v_{23}$,\\
&& $S_0$[3][49+t]=$v_{24}$,$S_0$[3][50+t]=$v_{25}$,$S_0$[3][55+t]=$v_{26}$,$S_0$[3][56+t]=$v_{27}$,\\
&& $S_0$[3][58+t]=$v_{28}$,$S_0$[3][59+t]=$v_{29}$,$S_0$[3][62+t]=$v_{30}$,$S_0$[3][63+t]=$v_{31}$.\\
\cline{2-3}
&key
& If a cube sum is nonzero, $k_0(t)+ k_1(t)= 1$.\\
\cline{1-3}
\multirow{9}{*}{(4)}&
&$S_0$[3][t]=$v_{0}$,$S_0$[3][1+t]=$v_{1}$,$S_0$[3][3+t]=$v_{2}$,$S_0$[3][4+t]=$v_{3}$,\\
&& $S_0$[3][5+t]=$v_{4}$,$S_0$[3][6+t]=$v_{5}$,$S_0$[3][8+t]=$v_{6}$,$S_0$[3][14+t]=$v_{7}$,\\
&& $S_0$[3][15+t]=$v_{8}$,$S_0$[3][16+t]=$v_{9}$,$S_0$[3][17+t]=$v_{10}$,$S_0$[3][20+t]=$v_{11}$,\\
&& $S_0$[3][26+t]=$v_{12}$,$S_0$[3][27+t]=$v_{13}$,$S_0$[3][29+t]=$v_{14}$,$S_0$[3][30+t]=$v_{15}$,\\
&cube& $S_0$[3][33+t]=$v_{16}$,$S_0$[3][34+t]=$v_{17}$,$S_0$[3][35+t]=$v_{18}$,$S_0$[3][37+t]=$v_{19}$,\\
&& $S_0$[3][38+t]=$v_{20}$,$S_0$[3][39+t]=$v_{21}$,$S_0$[3][40+t]=$v_{22}$,$S_0$[3][46+t]=$v_{23}$,\\
&& $S_0$[3][49+t]=$v_{24}$,$S_0$[3][50+t]=$v_{25}$,$S_0$[3][55+t]=$v_{26}$,$S_0$[3][58+t]=$v_{27}$,\\
&& $S_0$[3][59+t]=$v_{28}$,$S_0$[3][60+t]=$v_{29}$,$S_0$[3][62+t]=$v_{30}$,$S_0$[3][63+t]=$v_{31}$.\\
\cline{2-3}
& nonce & $S_0$[4][t]=0.\\
\cline{2-3}
& key
& If a cube sum is nonzero, $k_0(t)+ k_1(t)=0$.\\
\hline
\multirow{10}{*}{(5)}&
& $S_0$[3][t]=$v_{0}$,$S_0$[3][1+t]=$v_{1}$,$S_0$[3][4+t]=$v_{2}$,$S_0$[3][5+t]=$v_{3}$,$S_0$[3][6+t]=$v_{4}$,\\
&& $S_0$[3][8+t]=$v_{5}$,$S_0$[3][9+t]=$S_0$[4][9+t]=$v_{6}$,$S_0$[3][11+t]=$S_0$[4][11+t]=$v_{7}$,\\
&& $S_0$[3][14+t]=$v_{8}$,$S_0$[3][15+t]=$v_{9}$,$S_0$[3][16+t]=$v_{10}$,\\
&& $S_0$[3][18+t]=$S_0$[4][18+t]=$v_{11}$,$S_0$[3][24+t]=$S_0$[4][24+t]=$v_{12}$,\\
&&$S_0$[3][26+t]=$v_{13}$,$S_0$[3][27+t]=$v_{14}$,$S_0$[3][29+t]=$v_{15}$,$S_0$[3][30+t]=$v_{16}$,\\
&cube&$S_0$[3][34+t]=$v_{17}$,$S_0$[3][36+t]=$v_{18}$,$S_0$[3][37+t]=$v_{19}$,$S_0$[3][38+t]=$v_{20}$,\\
&& $S_0$[3][39+t]=$v_{21}$,$S_0$[3][45+t]=$v_{22}$,$S_0$[3][47+t]=$S_0$[4][47+t]=$v_{23}$,\\
&& $S_0$[3][48+t]=$v_{24}$,$S_0$[3][49+t]=$v_{25}$,$S_0$[3][50+t]=$v_{26}$,$S_0$[3][56+t]=$v_{27}$,\\
&& $S_0$[3][58+t]=$v_{28}$,$S_0$[3][59+t]=$v_{29}$,$S_0$[3][60+t]=$v_{30}$,$S_0$[3][63+t]=$v_{31}$.\\
\cline{2-3}
&nonce & $S_0$[4][t]=0,$S_0$[4][4+t]=0,$S_0$[4][16+t]=0,$S_0$[4][58+t]=0,$S_0$[4][63+t]=0.\\
\cline{2-3}
& key
& If a cube sum is nonzero, $k_0(t)+ k_1(t)=0$.\\
\hline
\noalign{\smallskip}
\end{tabular}
\caption{Parameters Set for Attack on the 6-round Initialization of \textsc{Ascon}}
\label{tab:6rattack}
\end{table}

\section{Test 2}\label{app:test2}
\subsection{Test 2 for 5-round initialization of \textsc{Ascon}-128}
We test 5-round cube sums over a 17-dimension random cube with 1000 keys. The random cube set in two-word nonce is as Figure~\ref{fig:5r17dim}, where each grey bit represents one cube variable in the two-word nonce. For each bit of the 64-bit output of 5-round initialization of \textsc{Ascon}-128, the accumulation and the ratio of nonzero cube sums among 1000 keys are listed in Table~\ref{tab:5r17dim1000keys}.
\begin{center}
\includegraphics[height=0.42cm]{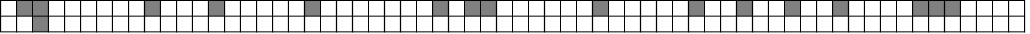}
\makeatletter\def\@captype{figure}\makeatother
\caption{The Random Cube Set with Dimension 17}
\label{fig:5r17dim}
\end{center}

\begin{table}
\caption{Statistics of Nonzero Cube Sums in the 64-bit Output of 5-round \textsc{Ascon}-128}
\label{tab:5r17dim1000keys}
\begin{center}
\begin{tabular}{|c|c|c|c|c|c|}
\hline
bit&\#nonzero&\%nonzero&bit&\#nonzero&\%nonzero\\
\hline
0&0&0&32&125&0.125                           \\
\hline
1&268&0.268&33&110&0.11                      \\
\hline
2&0&0&34&0&0                                 \\
\hline
3&0&0&35&6&0.006                             \\
\hline
4&198&0.198&36&37&0.037                     \\
\hline
5&91&0.091&37&350&0.35                      \\
\hline
6&75&0.075&38&0&0                           \\
\hline
7&14&0.014&39&92&0.092                     \\
\hline
8&31&0.031&40&282&0.282                     \\
\hline
9&92&0.092&41&0&0                           \\
\hline
10&74&0.074&42&0&0                         \\
\hline
11&16&0.016&43&20&0.02                    \\
\hline
12&123&0.123&44&0&0                         \\
\hline
13&0&0&45&6&0.006                           \\
\hline
14&0&0&46&106&0.106                         \\
\hline
15&75&0.075&47&0&0                         \\
\hline
16&0&0&48&92&0.092                         \\
\hline
17&22&0.022&49&4&0.004                     \\
\hline
18&86&0.086&50&61&0.061                   \\
\hline
19&0&0&51&75&0.075                         \\
\hline
20&92&0.092&52&31&0.031                   \\
\hline
21&4&0.004&53&0&0                           \\
\hline
22&0&0&54&0&0                               \\
\hline
23&125&0.125&55&0&0                         \\
\hline
24&95&0.095&56&272&0.272                   \\
\hline
25&0&0&57&0&0                               \\
\hline
26&26&0.026&58&0&0                         \\
\hline
27&31&0.031&59&156&0.156                   \\
\hline
28&92&0.092&60&0&0                         \\
\hline
29&0&0&61&19&0.019                         \\
\hline
30&16&0.016&62&28&0.028                   \\
\hline
31&134&0.134&63&0&0                         \\
\hline
\end{tabular}
\end{center}
\end{table}

\subsection{Test 2 for 6-round initialization of \textsc{Ascon}-128}
We test 6-round cube sums over a 33-dimension random cube with 982 keys. The random cube set in two-word nonce is as Figure~\ref{fig:6r33dim}, where each grey bit represents one cube variable in the two-word nonce. For each bit of the 64-bit output of 6-round initialization of \textsc{Ascon}-128, the accumulation and the ratio of nonzero cube sums among 982 keys are listed in Table~\ref{tab:6r33dim982keys}.

\begin{center}
\includegraphics[height=0.42cm]{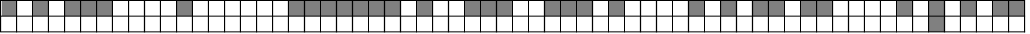}
\makeatletter\def\@captype{figure}\makeatother
\caption{The Random Cube Set with Dimension 33}
\label{fig:6r33dim}
\end{center}

\begin{table}
\caption{Statistics of Nonzero Cube Sums in the 64-bit Output of 6-round \textsc{Ascon}-128}
\label{tab:6r33dim982keys}
\begin{center}
\begin{tabular}{|c|c|c|c|c|c|}
\hline
bit&\#nonzero&\%nonzero&bit&\#nonzero&\%nonzero\\
\hline
0&512&0.521385&32&488&0.496945              \\
\hline
1&528&0.537678&33&483&0.491853              \\
\hline
2&494&0.503055&34&506&0.515275              \\
\hline
3&489&0.497963&35&487&0.495927              \\
\hline
4&501&0.510183&36&503&0.51222               \\
\hline
5&499&0.508147&37&488&0.496945              \\
\hline
6&481&0.489817&38&487&0.495927              \\
\hline
7&492&0.501018&39&499&0.508147              \\
\hline
8&472&0.480652&40&491&0.5                   \\
\hline
9&501&0.510183&41&479&0.48778               \\
\hline
10&472&0.480652&42&483&0.491853            \\
\hline
11&479&0.48778&43&465&0.473523              \\
\hline
12&504&0.513238&44&494&0.503055            \\
\hline
13&491&0.5&45&478&0.486762                  \\
\hline
14&511&0.520367&46&492&0.501018            \\
\hline
15&500&0.509165&47&474&0.482688            \\
\hline
16&474&0.482688&48&499&0.508147            \\
\hline
17&485&0.49389&49&479&0.48778               \\
\hline
18&484&0.492872&50&508&0.517312            \\
\hline
19&482&0.490835&51&486&0.494908            \\
\hline
20&469&0.477597&52&479&0.48778             \\
\hline
21&491&0.5&53&500&0.509165                  \\
\hline
22&492&0.501018&54&487&0.495927            \\
\hline
23&493&0.502037&55&496&0.505092            \\
\hline
24&477&0.485743&56&495&0.504073            \\
\hline
25&519&0.528513&57&477&0.485743            \\
\hline
26&498&0.507128&58&497&0.50611             \\
\hline
27&507&0.516293&59&489&0.497963            \\
\hline
28&484&0.492872&60&494&0.503055            \\
\hline
29&501&0.510183&61&494&0.503055            \\
\hline
30&493&0.502037&62&484&0.492872            \\
\hline
31&482&0.490835&63&483&0.491853\\
\hline
\end{tabular}
\end{center}
\end{table}
\section{Attacks on 7-round initialization of \textsc{Ascon}-128a}\label{app:ascon128a}
The complexity of the attack on 7-round initialization of \textsc{Ascon}-128a is analysed carefully here. With 0 refers to the most significant bit, in \textsc{Ascon}-128 six bits of $IV$ value 1, whose indexes are $\{0,9,20,21,29,30\}$, while it turns to 5 bits for \textsc{Ascon}-128a, whose indexes are $\{0,8,20,21,28\}$. We compute the complexity of the worst case similar to \textsc{Ascon}-128. The range of the \emph{remaining key subset} is analysed as follow.
We denote the index set of key bits certainly determined by \textbf{CASE 2} as
\begin{equation}\label{eq.aki0}
KI_0=\{4,12,24,25,32,53,61,13,10,17\}
\end{equation}
For each of $( k_0(i), k_1(i) ) (i\in KI_0)$, only 2 guesses out of $2^2$ values of $(k_0(i), k_1(i))$ obey the corresponding one of Eq.~(\ref{eq:controlkeydis1}) with $KI_0$ in Eq. (\ref{eq.aki0}). Each filtration ratio is $\frac{1}{2}$, so 10 $(k_0(i), k_1(i))$ lead a ratio $2^{-10}$. Additionally, the index sets of key bits certainly influenced by \textbf{CASE 1} is as follow:
\begin{eqnarray}\label{eq.aki1-ki3}
\left\{
\begin{aligned}
KI_{1}=\{&30,35,40,45,50,55,60,1,6,11,16,21,26,31,36,41,46,51,56\}\\
KI_{2}=\{&2,7\}\\
KI_{3}=\{&9,14,19\}\\
KI_{4}=\{&15,20\}\\
KI_{5}=\{&28,33,38,43,48\}\\
KI_{6}=\{&29,34,39,44,49,54,59,0,5\}\\
KI_{7}=\{&37,42,47,52,57,62,3,8\}\\
KI_{8}=\{&58,63\}\\
\end{aligned}
\right.
\end{eqnarray}
As shown in Eq. (\ref{eq.aki0}) and (\ref{eq.aki1-ki3}), $KI_j$ doesn't intersect with each other for $j\in\{0,1,...,8\}$. In each of $KI_j(j\in\{1,...,8\})$, suppose that $i_0$ is an arbitrary index and $i_1$ is next to it, then Eq. (\ref{eq.ai0i1}) should be hold.
\begin{equation}\label{eq.ai0i1}
 (k_0(i_0)+k_1(i_0)+1)(k_0(i_1)+k_1(i_1)+1)=0
\end{equation}
For $KI_j(j\in\{1,...,8\})$, the filtration ratios are $\frac{10946}{2^{19}}, \frac{3}{2^{2}}, \frac{5}{2^3}, \frac{3}{2^{2}}, \frac{13}{2^{5}}, \frac{89}{2^{9}},  \frac{55}{2^{8}}, \frac{3}{2^{2}}$ respectively. Due to the disappearance of key in any $U_i$ of \textbf{CASE 1}, it provides a filtration ratio equals $2^{-14.55}$ which is the product of the ratios above. The keys passed all the above testers will be checked by a $(nonce,plaintext,ciphertext,tag)$ pair to determine the right one.

The disappearance of key in any $U_i$ and $U'_i$ give a total filtration ratio of $2^{-10}\times2^{-14.55}$, and the keys with key bits passed through the testers above are exactly keys in the \emph{remaining key subset} whose size is $2^{128-10}\times 2^{-14.55}=2^{103.45}$. So in worst case, omitting complexity of cube computation, the total complexity is $2^{103.45}$.

\end{document}